\documentclass{aa}

\usepackage{txfonts}
\usepackage{graphicx,amsmath,amssymb}
\usepackage{longtable,lscape}
\usepackage[table]{xcolor}
\usepackage{natbib}

\bibpunct{(}{)}{;}{a}{}{,}

\usepackage{tikz}
\usepackage{pgfplots}

\usepackage[colorlinks=true,citecolor=blue,urlcolor=blue,linkcolor=blue]{hyperref}

\newcommand{\msun}{\ifmmode{{\rm M}_{\odot}}\else{${\rm M}_{\odot}$~}\fi}
\newcommand{\overbar}[1]{\mkern 2.5mu\overline{\mkern-2.5mu#1\mkern-2.5mu}\mkern 2.5mu}

\begin{document}

\title{
A grid of one-dimensional low-mass star formation collapse models
}

\author{
  N. Vaytet \& T. Haugb{\o}lle
}

\authorrunning{N. Vaytet \& T. Haugb{\o}lle}
\titlerunning{A grid of 1D low-mass star formation collapse models}

\institute{
  Centre for Star and Planet Formation, Niels Bohr Institute and Natural History Museum of Denmark, University of Copenhagen, {\O}ster Voldgade 5-7, DK-1350 Copenhagen K, Denmark\\ \email{neil.vaytet@nbi.ku.dk} 
}

\date{Received XX / Accepted XX}

\abstract
{Numerical simulations of star formation are becoming 
ever more sophisticated, incorporating new physical processes in increasingly realistic set-ups. These models are being compared to the latest observations through
state-of-the-art synthetic renderings that trace the different chemical species present in the protostellar systems. The chemical evolution of the interstellar and 
protostellar matter is very topical, with more and more chemical databases and reaction solvers available online to the community.}
{The current study was developed to provide a database of relatively simple numerical simulations of protostellar collapse as a template library for observations of cores and very young protostars, and for researchers who wish to test their 
chemical modelling under dynamic astrophysical conditions. It was also designed to identify statistical trends that may appear when running many models of the formation of 
low-mass stars by varying the initial conditions.}
{A large set of 143 calculations of the gravitational collapse of an isolated sphere of gas with uniform temperature and a Bonnor-Ebert-like density profile was undertaken using a 1D fully implicit 
Lagrangian radiation hydrodynamics code. The parameter space covered initial masses from 0.2 to 8 M$_{\odot}$, temperatures of 5-30 K, and radii $3000 \le 
R_{0} \le 30,000$ AU.}
{A spread due to differing initial conditions and optical depths, was found in the thermal evolutionary tracks of the runs. Within less than an order of magnitude, all first and second Larson cores had masses and radii essentially independent of the initial conditions. Radial profiles of the gas density, velocity, and temperature were found to vary much more outside of the first core than inside.
The time elapsed between the formation of the first and second cores was found to strongly depend on the first core mass accretion rate, and no first core in our grid of models lived for longer than 2000 years before the onset of the second collapse.}
{The end product of a protostellar cloud collapse, the second Larson core, is at birth a canonical object with a mass and radius of about 3 M$_{\mathrm{J}}$ and 8 R$_{\mathrm{J}}$, independent of its initial conditions. The evolution sequence which brings the gas to stellar densities can, however, proceed in a variety of scenarios, on different timescales or along different isentropes, but each story line can largely be predicted by the initial conditions. All the data from the simulations are publicly available.\thanks{The figures and raw data for every simulation output can be found at this address: \url{http://starformation.hpc.ku.dk/grid-of-protostars}. Copies of the outputs, as well as Table~\ref{tab:parameters}, are also available in the form of static electronic tables at the CDS via anonymous ftp to \url{cdsarc.u-strasbg.fr} (130.79.128.5) or via \url{http://cdsweb.u-strasbg.fr/cgi-bin/qcat?J/A+A/}.}}


\maketitle


\section{Introduction}\label{sec:intro}

The generally accepted view that stars form by gravitational condensation of diffuse interstellar matter is very old; however, it is only in
recent decades that we have gained physical understanding of the process,
thanks to ever more precise observations and complex numerical models \citep{MckeeOstriker2007,AndreEtAl2014}.
Stars form within turbulent molecular clouds when density fluctuations are large enough to become gravitationally unstable and begin to collapse onto themselves.
The details of this process remain largely unclear, as many different physical mechanisms operate over a very wide range of spatial scales.
The progenitorial density enhancements (also known as parent clouds) are initially optically thin and contract isothermally, while the compression heating is 
radiated away. As the density of the gas inside the cloud rises, the efficiency of the radiative cooling drops until it can no longer counter-balance the compressive 
calefaction. The gas evolution becomes adiabatic, and a hydrostatic body, known as the first Larson core after the pioneering numerical studies of \citet{Larson1969}, 
is formed at the centre of the collapsing system. It accretes material from its surrounding envelope, and the sustained increase in mass, density, and temperature 
eventually triggers the dissociation of $\text{H}_{2}$ molecules, when temperatures exceed $\sim$2000 K. This leads to the second phase of collapse
because of the endothermic nature of the dissociation process. The collapse ceases when most or all of the $\text{H}_{2}$ molecules
have been split, at which point a much more dense and compact hydrostatic core (Larson's second core) is formed \citep{Larson1969,Masunaga2000,Vaytet2013a}.
The temperature inside the second core continues to rise until the ignition of nuclear reactions: the young star is born.

This multi-scale, multi-physics problem is very difficult to tackle numerically;
the parent cloud scales $\sim$10$^{4}$ astronomical units (AU), while the protostar measures only 10\textsuperscript{-3} AU at birth.
An intricate interplay operates between large-scale environmental factors, which regulate the 
supply of mass, angular momentum, and
magnetic flux \citep{Li2015}, and small-scale processes close to the star,
which control the evolution and dynamics of protostellar systems \citep{HennebelleCharbonnel2013}.
Numerical models are becoming more sophisticated every year, incorporating new physical processes 
\citep{Commercon2010,FederrathSchron2014,BateTricco2014,Tomida2015,Gonzalez2015,Masson2016,Hopkins2016,Wurster2016}
in increasingly realistic set-ups
\citep{Krumholz2012,Seifried2012,Li2014,NordlundHaugbolle2014,Padoan2014,Federrath2015,Matsumoto2015}\footnote{This list is by no means exhaustive, it simply reflects some of the recent developments in computational star formation.}.
The comparison between these models and the latest observations
has also reached new heights in recent years, using radiative transfer tools and observatory simulators to produce astonishing synthetic maps and spectra 
\citep{Commercon2012a,Commercon2012b,FrimannHaugbolle2016,FrimannJorgensen2016,Visser2015,Kataoka2015,Seifried2016}.
This analysis is heavily focused on the emission from the dust and
the different chemical species present in the
protostellar systems, and determining which structures they trace (discs, outflows, protostar). While the chemical abundances in these studies are derived chiefly from 
the dust and/or gas temperatures, much effort is currently devoted to including time-dependent non-equilibrium chemical evolution in such simulations 
\citep{GrassiBovino2014,Kim2014,Hincelin2016,DzyurkevichCommercon2016,Grassi2016}. Comprehensive databases of interstellar chemical reaction rates, such as the \textsc{umist} database \citep{McElroy2013} or \textsc{kida} \citep{Wakelam2015} for 
instance, are now also publicly available to the community.

In this context, the current study was developed to provide a database of relatively simple numerical simulations of protostellar collapse for researchers who wish to 
test their chemical modelling under dynamic astrophysical conditions. Instead of producing a collection of large complex turbulent 3D simulations, 
which have a prohibitively high computational cost and can be very cumbersome to analyse and interpret, the aim was to use a simple spherically symmetric set-up and varying the initial parameters. The study extends the 
simulation set in \citet{Vaytet2013a} with 143 new calculations of the gravitational collapse of an isolated sphere of gas with uniform temperature and a Bonnor-Ebert-like profile. The 
calculations solve the equations of radiation hydrodynamics with self gravity, using a non-ideal gas equation of state and an up-to-date table of dust and gas opacities. 
The code uses a Lagrangian mesh which follows the evolution of each piece of fluid throughout the entire simulation, making it ideal for tracing the chemical evolution of 
gas elements. This paper describes the numerical method used and the initial parameters for the different runs, and also reports on statistical trends that 
emerge from running a large set of collapse models. The masses and radii of the first and second Larson cores, among others, are given and the connection between the 
different initial parameters and final core properties are investigated.


\section{Numerical method and initial conditions} \label{sec:num}

The \texttt{SINERGHY1D} code was used to solve the equations of radiation hydrodynamics; it is a 1D fully implicit Godunov Lagrangian code, described in 
\citet{Vaytet2012,Vaytet2013a}.
It uses the $M_{1}$ closure to model the radiative transfer \citep{Levermore1984,Dubroca1999} and the grid comprises 4096 cells (see Appendix~\ref{app:resolution}). The 
mesh is fully Lagrangian, no re-gridding scheme is employed in this work, contrary to what has been done in some of our previous studies.
To have a higher resolution towards the centre of the mesh, the cell size is increased progressively towards larger radii. The size of cell $i+1$ is defined by $dr_{i+1} = (1+\alpha)~dr_{i}$, where we have chosen $\alpha = 8 \times 10^{-4}$.
The integration timestep was chosen so that the variations from one time step to the next of each primary hydrodynamic (mesh position, velocity and total energy) and 
radiative (energy density and flux) variable was limited to a maximum of 10\%. This ensured rapid convergence of the Newton-Raphson iterations in our time-implicit 
scheme.

The code incorporates the gas equation of state of \citet{Saumon1995}, and its extension to low densities \citep[see][]{Vaytet2013a}, for a mixture of Hydrogen and Helium,
with a Helium mass concentration of 0.27.
The interstellar dust and gas opacities were re-used from \citet{Vaytet2013a}.
These comprise, at low temperatures (below 1500 K), the opacities of 
\citet{Semenov2003} as the absorption coefficients are dominated by the one percent in mass of dust grains. For temperatures between $\sim$1500-3200 K, the dust grains are rapidly destroyed and molecular gas opacities prevail. These were obtained 
from the model of \citet{Ferguson2005}. Finally, at temperatures above $\sim$3200 K, there are no more molecules and the atomic opacities from the OP project 
\citep{Badnell2005} were employed. We also made use of the opacity averaging scheme presented in \citet{Vaytet2013b} to transition between Planck and Rosseland opacity in regions of different optical thickness.

\input{epsilonBE.tex}

The initial set-up for the dense cloud collapse was slightly modified compared to \citet{Vaytet2013a}, and somewhat similar to \citet{Tomida2013}. A Bonnor-Ebert \citep{Bonnor1956,Ebert1955} like density profile was adopted, with a density contrast between the centre and the edge of the sphere $\rho_{\mathrm{c}}/\rho_{0} = 14.1$, corresponding to a dimensionless radius of $\xi = 6.45$. To fully define the gas density, fixed values for the sphere's temperature $T_{0}$ and radius $R_{0}$ were chosen. The density at the centre of the cloud for a critical Bonnor-Ebert (BE) sphere is then found via
\begin{equation}\label{equ:rhoc_crit}
\rho_{\mathrm{c},\text{crit}} = \left(\frac{\xi}{R_{0}}\right)^{2} \frac{k_{\mathrm{B}}T_{0}}{4\pi \mu m_{\mathrm{H}} G} ~,
\end{equation}
where $k_{\text{B}}$ is Boltzmann's constant, $\mu$ represents the mean atomic weight $(=2.31)$, $m_{\text{H}}$ the hydrogen atom mass, and $G$ the gravitational constant.
This is used to scale the dimensionless BE density profile to physical densities. Integrating the mass inside the cloud gives the same result as the well-known critical BE mass
\begin{equation}\label{equ:M_BE}
M_{\mathrm{BE}} = 2.4 \frac{c_{0}^{2}}{G} R_{0} ~,
\end{equation}
where $c_{0}$ is the isothermal sound speed of the gas.
Finally, to obtain a given initial cloud mass $M_{0}$, we divide the density everywhere in the cloud by the ratio of BE mass to cloud mass $\epsilon = M_{\text{BE}}/M_{0}$, which determines whether a cloud is stable ($\epsilon>1$) or unstable ($\epsilon<1$) to gravitational collapse.
Combining equations~(\ref{equ:rhoc_crit}) and (\ref{equ:M_BE}) with the definition of $\epsilon$, we see that the central density is independent of the temperature, only depending on the average density in the cloud, and is given by
\begin{equation}\label{equ:rhoc_cloud}
\rho_{\mathrm{c}} = 5.78~M_{0} \left(\frac{4 \pi}{3} R_{0}^{3}\right)^{-1} ~.
\end{equation}
The cloud's free-fall time, defined as
\begin{equation}\label{equ:tff}
t_{\text{ff}} = \sqrt{ \frac{3\pi}{32 G \rho_{\mathrm{c}}} }  ~,
\end{equation}
gives a good estimate of the time needed to form the first Larson core. The formation of the second Larson core will
follow almost `instantaneously' (on a timescale of $\sim$100-1000 years).
The radiation temperature is initially in equilibrium with the gas temperature and the radiative flux is set to zero
everywhere. The boundary conditions are reflexive at the centre of the grid ($r = 0$). At the outer edge of the sphere, the hydrodynamical variables have a zero-gradient 
boundary condition, while the radiative variables are
fixed to their initial values inside the ghost cells. The equations of radiative transfer were integrated over all frequencies (grey approximation) since including
frequency dependence only yields small differences for a much increased computational cost \citep{Vaytet2012,Vaytet2013a}.

The parameter space was chosen to cover a large range of $\epsilon$ between 0 and 1, with different initial cloud masses, temperatures and radii.
The cloud masses range from 0.2~\msun to $8~\msun$, the temperatures cover $5-30$ K, while six separate radii were chosen between 3000 and 30,000 AU.
The parameter space is illustrated in Fig.~\ref{fig:epsilons}, and all the run parameters are listed in Table~\ref{tab:parameters}.
Only gravitationally unstable clouds ($\epsilon < 1$) were selected. The dividing line between stable and unstable clouds, for a given radius is defined by
\begin{equation}\label{equ:epseq1}
T_{0} = \frac{\xi}{C_{\text{BE}}}  \left(\frac{\mu m_{\text{H}}G}{k_{\text{B}}}\right)  \frac{M_{0}}{R_{0} \sqrt{4\pi \rho_{\text{c}}/\rho_{0}}} ~, 
\end{equation}
where $C_{\text{BE}}\simeq 1.18$ is a constant.
Each simulation was stopped shortly after the formation of the second Larson core. We were not able to follow the evolution of the system for very long beyond this point, 
even with our time-implicit scheme. The integration timesteps are limited to ensure small enough relative variations (<10\%) of the hydrodynamic and radiative variables 
from one time step to the next. Having a Lagrangian mesh with two more or less stationary accretion shocks implies that pieces of fluid must travel through the 
shocks, and as they do, they suffer large increases in density, temperature and radiative flux. This has the consequence of keeping the integration timestep very low 
(especially after the second core has formed, as it has very high accretion velocities), making any lengthy integrations into the main accretion phase very costly. Other 
authors \citep[][for instance]{Foster1993,Masunaga2000} have commonly switched to a Eulerian scheme once the core is formed, to follow the system's evolution on much 
longer timescales. Re-writing an Eulerian version of our code to simulate the main accretion phase is deferred to future work.


\section{Results}\label{sec:results}

For the details on the evolution of a typical run, which forms a first and second Larson core, the reader is refered to the previous paper by \citet{Vaytet2013a} which 
provides an extensive discussion on the evolution and structure of spherically symmetric collapsing clouds. In the present work, we shall limit ourselves to studying 
trends in the results from the multiple simulations performed, attempting to extract statistically meaningful relationships between initial parameters and final core 
properties.


\subsection{Central densities, temperatures and entropies}\label{sec:thermal_evolution}

\begin{figure*}
\centering
\includegraphics[width=\textwidth]{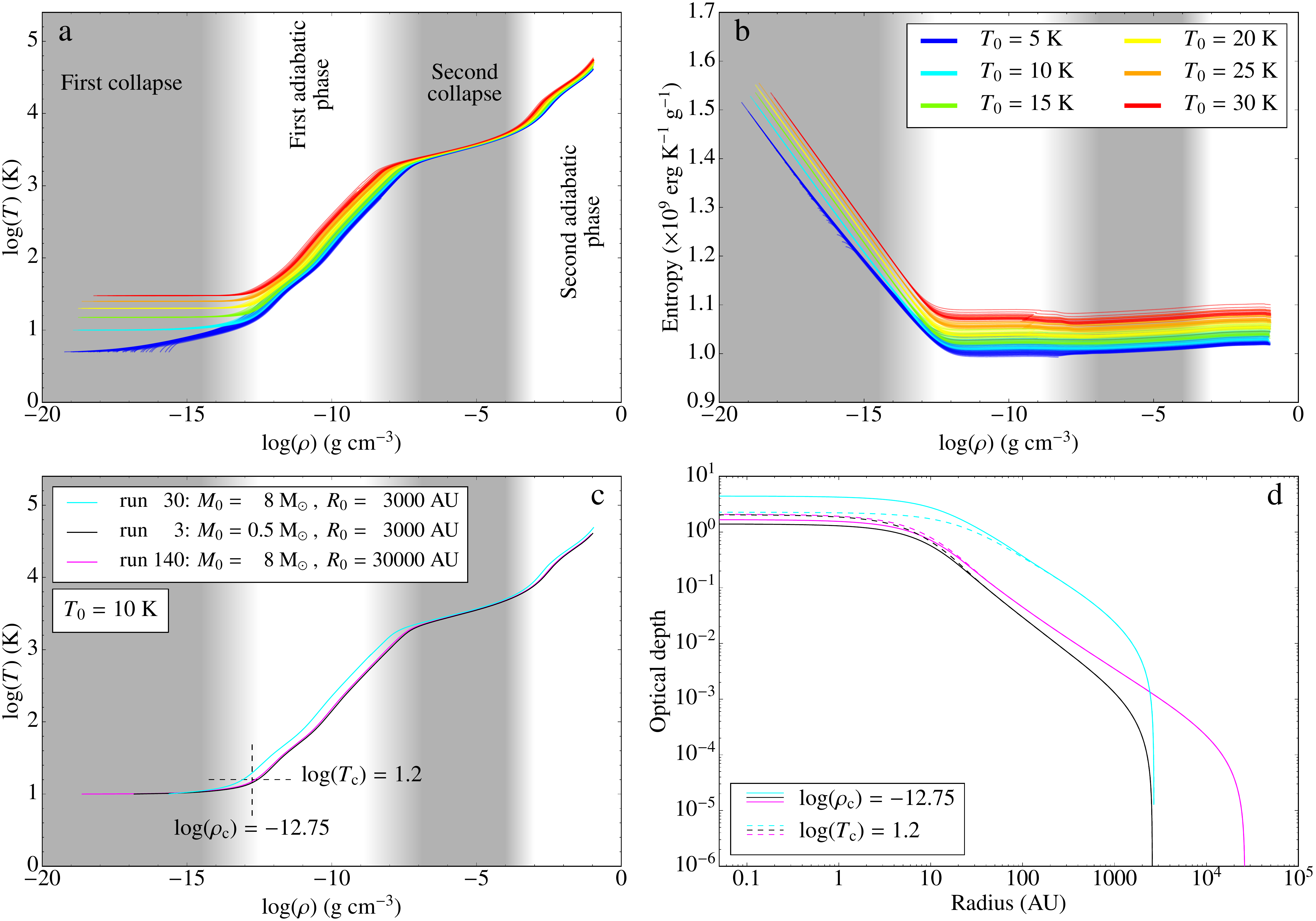}
\caption{\underline{\smash{Top row}}: evolution of the central temperature (a) and entropy (b) as a function of density for all 143 runs. The lines have been colour-coded according to the cloud's initial temperature $T_{0}$ (see legend in panel b).
\underline{\smash{Bottom row}}: (c) Evolution of the central temperature as a function of density for runs 30, 3, and 140 (see legend). (d) Optical depth of the systems as a function of radius for runs 30, 3, and 140 (same colours as in panel c), 
when the central density has just reached $\log(\rho_{\mathrm{c}}) =-12.75~\text{g~cm}^{-3}$ (solid), and when the central temperature has reached $\log(T_{\mathrm{c}}) = 1.2$ K (dashed).}
\label{fig:all_rhoT}
\end{figure*}

Figure~\ref{fig:all_rhoT}a shows the thermal evolution (temperature as a function of density) for the central cell in the computational grid. All the tracks follow a 
similar evolution. They first go through an initial isothermal phase, where the low optical depth allows compression heating to be radiated away. As the density rises, 
the efficiency of the radiative cooling drops and the system begins to heat up, following an adiabat with constant entropy (see panel b). This 
adiabatic phase is refered to as the formation of the first Larson core.

\begin{figure*}
\centering
\includegraphics[width=\textwidth]{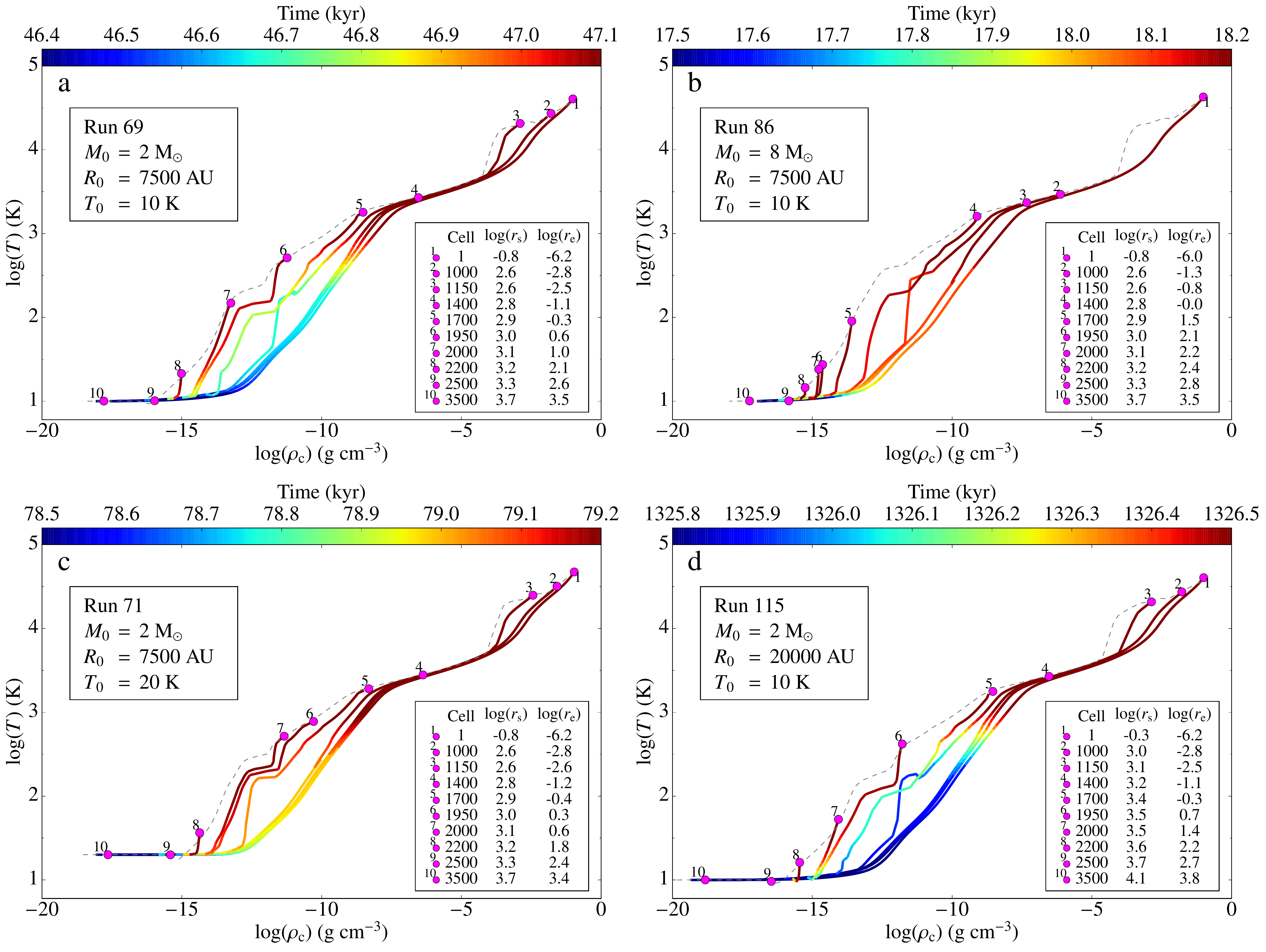}
\caption{Evolution track in the density-temperature plane for 10 different pieces of fluid in runs 69 (a), 86 (b), 71 (c), and 115 (d). In all panels, 10 cells were 
selected and followed throughout the simulation. The number of the cell in the mesh is given for each track in the bottom-right inset, along with the cell's radius at the 
start of the simulation (`s') and at the end (`e'). While the cell numbers are the same in all panels, the cell radii change. The colour codes for the time in kyr; the 
colourbar is different in all panels, but they all display the final 700 years of evolution. The pink dots show the position of each piece of fluid at the end of the simulation. The dashed grey line traces the final $(\rho,T)$ 
profile in the entire grid, which connects the pink dots. The inset in the top left corners gives a reminder of the parameters of each run.}
\label{fig:fluid_tracks}
\end{figure*}

The 143 runs show a wide spread in the thermal tracks they follow during this first adiabatic phase; its origin is two-fold.
First, as shown from the colour coding in the top row of Fig.~\ref{fig:all_rhoT}, a vertical spread is seen to be the direct consequence of the different initial temperatures for the various runs. In the first adiabatic phase, the effective ratio of specific heats of the gas $\gamma_{\text{eff}}$ changes from 1 (isothermal collapse) to $\sim$7/5. This sets the slope of the tracks, which is approximately the same for all runs. This implies that a system starting with a higher initial temperature $T_{0}$, as it becomes optically thick, will follow an adiabatic track that is shifted upwards in the vertical direction.
Moreover, the Rosseland mean opacity of the dust in the low temperature region behaves like $\kappa_{\text{R}} \propto T^{2}$. The runs with high initial temperatures will therefore have higher optical depths even when the density is the same, and will consequently become adiabatic earlier, attaining higher entropies (as seen in Fig~\ref{fig:all_rhoT}b).
In addition, a second spread is also observed in runs with identical $T_{0}$, this time in the horizontal direction.
To determine the origin of this spread, we select three runs (30, 3, and 140) with $T_{0} = 10$ K but different initial masses and radii. The thermal evolutions of these three runs is displayed in 
Fig.~\ref{fig:all_rhoT}c. Runs 30 and 3 begin along the same isothermal path, but start heating up at different central densities, following separate tracks in the 
first adiabatic phase. This first differentiation appears to largely determine the subsequent evolution of the system, as the runs never re-join the same tracks during 
the rest of the simulations. The cloud begins to heat up once the radiative cooling is no longer efficient compared to the compressional heating \citep[see the discussions 
in][]{Masunaga1998,Masunaga1999}. Figure~\ref{fig:all_rhoT}d shows the optical depth in the systems at two different times: when the central density has just reached $\log(\rho_{\mathrm{c}}) =-12.75~\text{g~cm}^{-3}$ (solid lines), and when the central temperature hits 16 K (dashed lines). It is clear that, for the chosen central density (when the spread in 
temperature is noticeable), the higher the initial mass, the higher the optical depth, and hence the more the cloud has heated up. The dashed curves 
illustrate that the runs show the same temperature rise for approximately the same optical depth. Comparing to run 140 supports further the statement that the evolution is governed by the optical depth. Run 140 has the same mass as run 30, but a radius 10 times larger, greatly diminishing its optical depth. In fact, it turns out to have an optical depth very similar to run 3 (see panel d), and both runs have matching evolutions.
In our case, isothermality breaks down around an optical depth of unity 
or slightly above, but we note that, as pointed out by \citet{Masunaga1999}, this does not by any means constitute a general rule. The determining factor is the balance 
between radiative cooling and compression heating, which is not always ruptured for an optical depth of unity. To estimate the density at which the cloud begins to depart from isothermality in its early evolution, we refer the reader to the exhaustive discussion in \citet{Masunaga1998}.

The first spread in thermal tracks at the start of the first adiabatic phase appear to also set the entropy level at the centre of the system for the remainder of its 
evolution. Indeed, Fig.~\ref{fig:all_rhoT}b shows that after a first phase of decrease during the isothermal collapse, the central entropy remains largely constant for $\rho_{c} > 10^{-12}~\text{g~cm}^{-3}$ (it should be noted that while the entropy at the centre of the system decreases, the entropy of the system as a whole increases). The spread in 
entropy at the centre of the protostar, at the end of the simulation, is quite significant compared to the average differential the central piece of fluid suffers during 
the entire simulation ($\sim$25\% of the average total differential). This implies that the different protostellar seeds formed in our simulation will have different 
early evolutions, as small differences in entropy can cause large variations in luminosity \citep[see for example][]{Stahler2005}.


\subsection{Evolutionary tracks for single pieces of fluid}\label{sec:fluid_tracks}

As this work was motivated by the evolution of chemical species in circumstellar matter going through various stages of protostellar evolution, we now turn to following 
the thermal evolution of other parts of the collapsing system, instead of just the centre. As many chemical evolution codes have no spatial dimensions and simply follow 
the chemical reactions and abundances in time, they could very easily use the thermal properties of a piece of fluid inside the mesh as a function of time as an input. 
And as our mesh is purely Langrangian, following a piece of fluid means selecting a single cell in the grid and tracing its time evolution. We plot in Fig.~\ref{fig:fluid_tracks}a the evolution in the $(\rho,T)$ plane of 10 different pieces of fluid in a `standard' run 69. It has an initial cloud mass of 2~\msun, temperature 
of 10~K, and an intermediate radius of 7500 AU. The 10 pieces of fluids were selected to illustrate different behaviours, and the 
cell number of each piece of fluid can be found in the lower-right inset in Fig.~\ref{fig:fluid_tracks}a. We see that the cells close to the centre of the system (tracks 
1 and 2) follow very closely the types of evolution shown in the previous section. Their temperatures increase smoothly as they are already beyond the first and second core 
accretion shocks when these are formed, and thus do not undergo rapid changes in density/temperature. Track 3 illustrates a piece of fluid that has just passed through the 
second core shock, showing a sharp rise in both density and temperature at late times. Tracks 4 and 5 both depict fluid parcels that have passed through the first core 
accretion shock, but have done so at different times.
In this diagram, a shock is identified by a discontinuity in density. Indeed, purely hydrodynamic shocks would also exhibit a temperature discontinuity, but strong 
supercritical radiative shocks (such as the first core accretion shock) have equal upstream and downstream temperatures. A shock is then identified because it implies 
strong compression, but not necessarily a jump in temperature.
Track 4 shows a sharp (horizontal) density jump around a density of $10^{-12}~\text{g~cm}^{-3}$, while the discontinuity in track 5 lies closer to $10^{-13}~\text{g~cm}^{-3}$. As we will see further down (Fig.~\ref{fig:cores_mass_radius}), the first core has a size of about 2~AU at the time of formation, which corresponds to the density 
of the jump in track 4 (this can be seen in the online data), and subsequently grows in size as it accretes material from the surrounding envelope. By the time track 5 
passes through the shock, the core's radius is located around 8~AU which now corresponds to a density of $10^{-13}~\text{g~cm}^{-3}$. In these two cases, the rises in 
temperature just upstream of the shock are due to radiative heating from the first core, which heats up the outer envelope. This is further illustrated by the final $(\rho,T)$ profile at the end of the simulation represented by the grey dashed line. Fluid 
parcel 6 has just gone through the accretion shock, whereas parcels 7 and 8 only get heated via radiation. It is important to note here, especially for chemical evolutions, that the 
inner parts of the system evolve much more rapidly that the outer envelope. Indeed, we can see that over the timeframe of the simulation, the inner parts reach 
protostellar densities and temperatures, while the outer gas stays below 100~K. This will inevitably lead to very contrasting chemical evolutions, depending on which 
piece of fluid is considered.

The evolution will also differ when different initial parameters are used. Figure~\ref{fig:fluid_tracks}b shows the evolution of pieces of fluid located at the same positions in the mesh, but for a 
simulation with a much smaller free-fall time (run 86 has the same initial radius and temperature than run 69, but four times the initial mass).
In this case, the most striking difference is that only the very central cells reach protostellar densities and temperatures. Tracks 2 and 3 have not yet reached the second core, while tracks 5 to 7 are still all outside of the first core at the end of the simulation.
The collapsing 
gas goes through the entire first adiabatic phase, second collapse and second adiabatic phase much faster than in run 69 (in run 86, most of the curve is coloured with shades of red and orange, meaning that this evolution takes place within the very last quarter of the time colour bar, for $t \geq 18$ kyr). This explains why the cells in the outer envelope 
remain at a low temperatures and densities in this run; the gas simply has 
not had enough time to travel inwards and heat up before the second core was formed and the simulation was stopped. As seen in panel (a), it does take some time for the temperature of fluid 
parcels 7 and 8 to rise.
Panel (c) illustrates the same evolution as in panel (a) but for run 71 that has an initial temperature of 20~K. While the outer cells, which sit at a temperature twice as high as in run 69, may see 
different chemical reactions dominating, the evolution of the rest of the system is very similar to run 69. We note, however, that the simulation end time has changed significantly in run 71 compared to run 69 due to the additional pressure support in the hotter cloud at the start of the simulation, showing that for marginally stable BE clouds, the free-fall time is only an order of magnitude estimate for the collapse time, as it does not consider pressure support in the cloud (runs 69 and 71 have identical free-fall times). Finally, panel (d) shows the fluid parcels' evolution in a 
case where the initial density is a factor of almost 20 lower. The free-fall time is about four times as long, and it will take the gas much longer to reach the high end of the 
$(\rho, T)$ space. As chemical reactions are highly time-dependent, this can also greatly modify the abundances of the species in the gas. Barring the different timescales, the evolution of the different fluid parcels are remarkably similar to those in run 69. Chemical abundances are of 
prime importance in protostellar evolution, as they regulate the gas hydrodynamics (through the heat capacity, line cooling, and chemical energy), the radiation (through the gas and dust opacities) and the 
magnetic field (through resistivities).


\subsection{Radial profiles}\label{sec:profiles}

\begin{figure*}
\centering
\includegraphics[width=\textwidth]{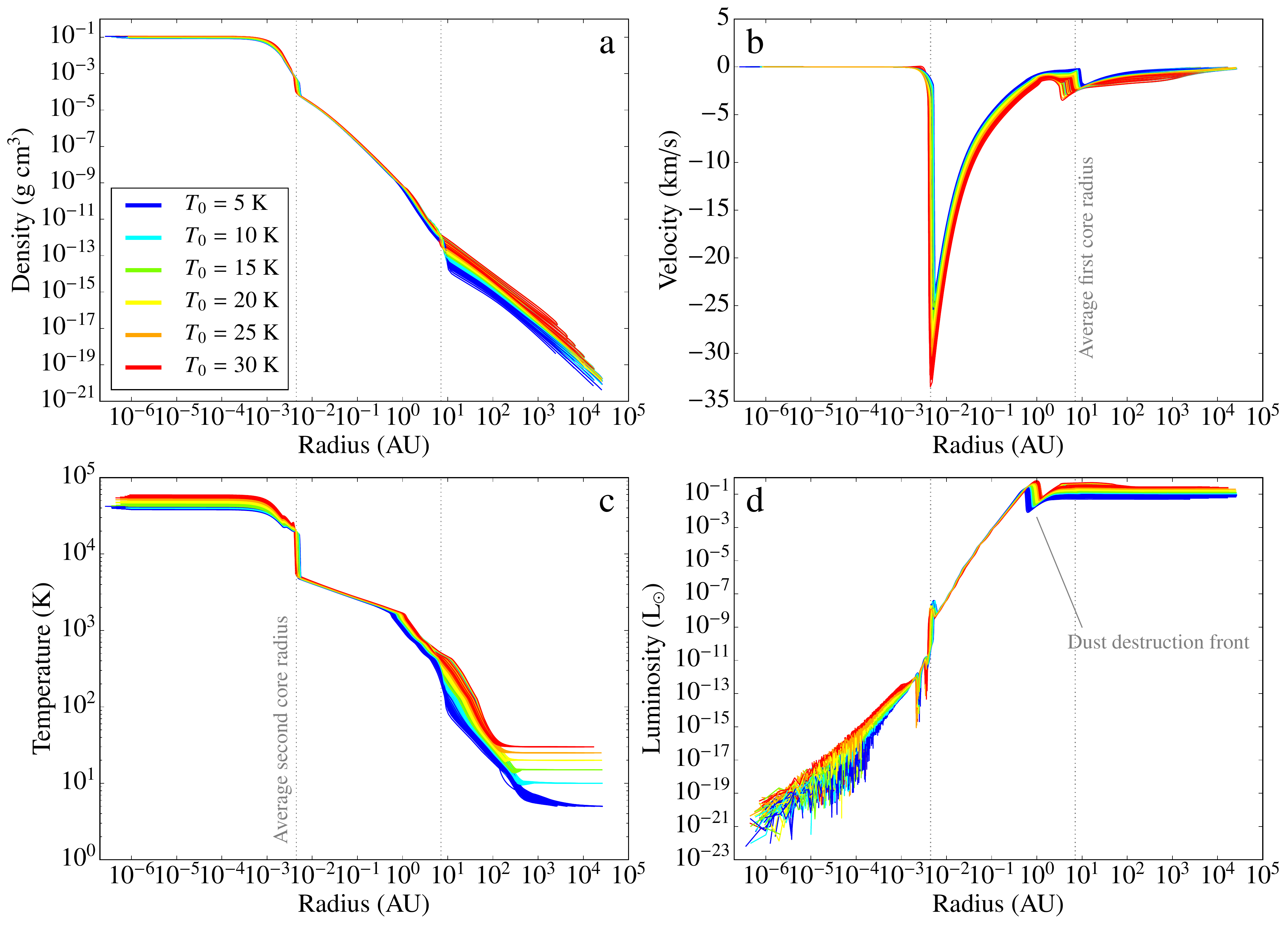}
\caption{Radial profiles of the gas density (a), velocity (b), temperature (c), and luminosity (d) for all 143 runs. The different colours represent different initial temperatures $T_{0}$ (see legend). The vertical dotted lines mark the positions of the average first and second core radii.}
\label{fig:profiles}
\end{figure*}

We now take a look at the radial profiles of various quantities and how they vary in the different runs. Figure~\ref{fig:profiles} shows the density, velocity, temperature and luminosity profiles for all 143 runs. As in Fig.~\ref{fig:all_rhoT}, the different colours represent different initial cloud temperatures. We first notice that, considering the range of initial parameters used, all the profiles are remarkably similar, especially inside the first core radius. The average first and second core radii are marked by the vertical grey dotted lines, and we see very little spread around these values. This result was already found in \citet{Vaytet2013a}, and the present study comfirms the statement accross a much larger variety of simulations. There is however some spread in the different profiles in the protostellar envelope, outside of the first core accretion shock. The initial densities and temperatures vary among the different runs, and it thus makes sense to see a spread in the outer regions of the system. This however shows that once a hydrostatic body is formed, the system `forgets' its initial conditions and all the profiles lie on top of each other. Finally, we find a spike in luminosity around 1 AU in all runs, which actually lies inside the first core and occurs at the dust destruction front, and not at the first core accretion shock. A small spike is also visible just at the second core accretion shock.


\subsection{Properties of the first and second Larson cores}\label{sec:core_props}

\begin{figure}
\centering
\includegraphics[width=0.48\textwidth]{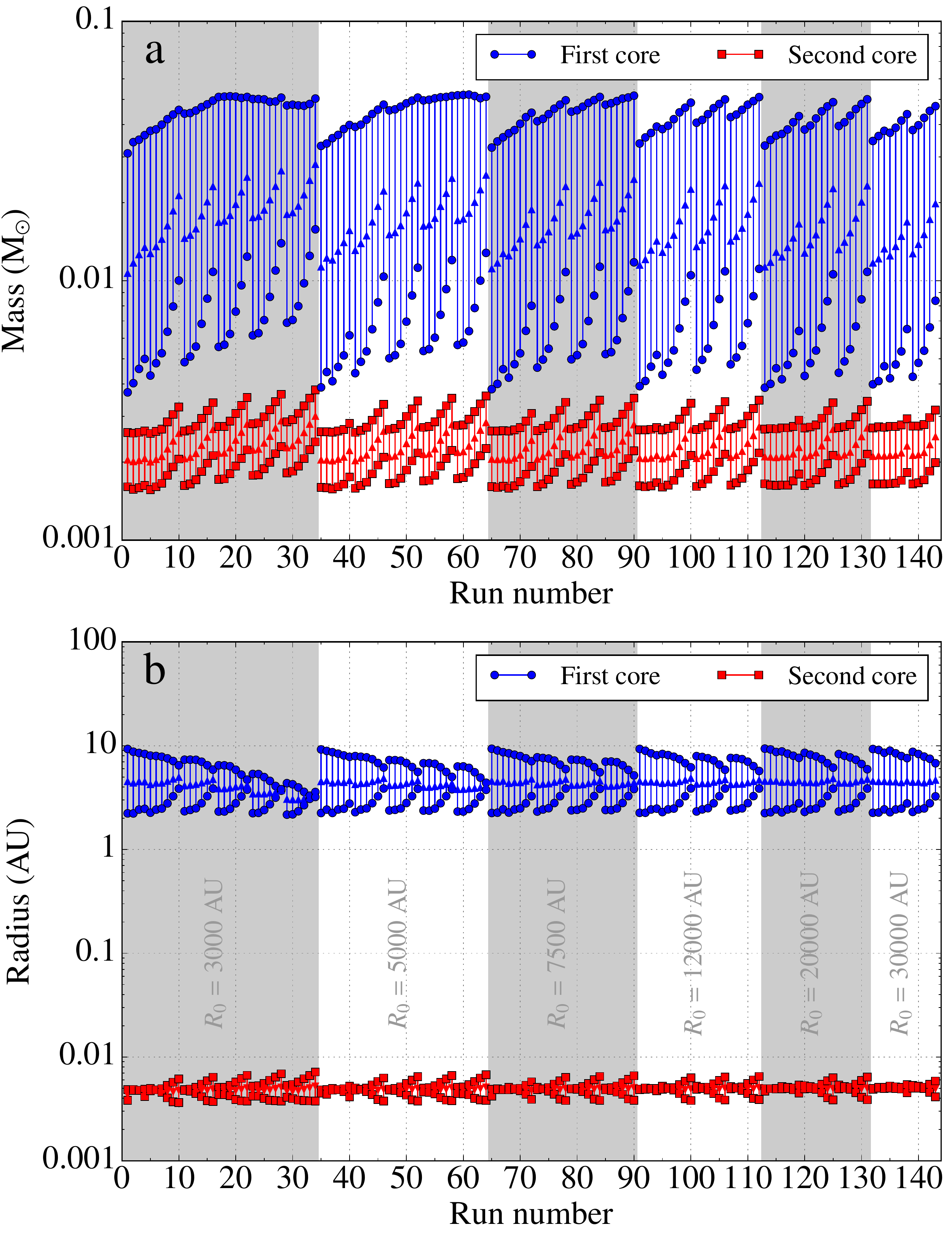}
\caption{First and second core mass (a) and radius (b) for all 143 runs. Each mass and radius has a starting and an end point. The starting point represents the quantity 
when the core has just formed, while the end point is at the end of the simulation. A small arrow in the middle of each segment points in the direction from start to end. 
The vertical alternating grey and white bands indicate changes in initial radius (see panel b for values of $R_{0}$).}
\label{fig:cores_mass_radius}
\end{figure}

Table~\ref{tab:parameters} lists the properties of the first and second Larson cores (mass, radius, etc.) at the end of each simulation. Figure~\ref{fig:cores_mass_radius} is a more graphical representation of masses (a) and radii (b) for the first (circles) and second (squares) cores.
It brings further information as it shows the cores’ masses and radii not only at the end of the simulation, but also just when the accretion shock at the core borders have formed.

\begin{figure*}
\centering
\includegraphics[width=\textwidth]{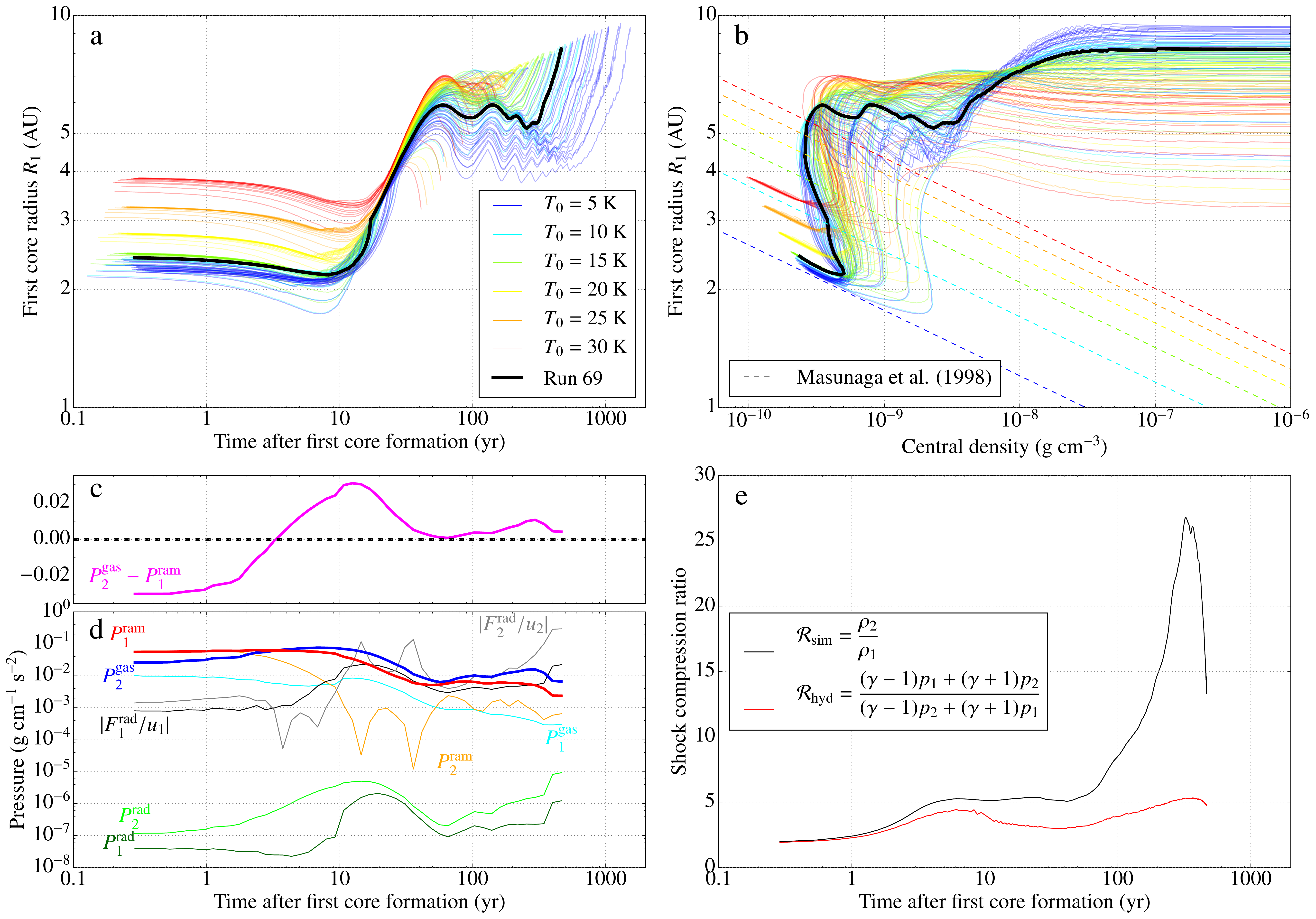}
\caption{(a) First core radius as a function of time, where $t = 0$ corresponds to the time when the first core accretion shock is formed. The different colours represent 
different initial cloud temperatures $T_{0}$ (see legend). The thick black line is for run 69. (b) First core radius as a function of central 
density (solid lines), alongside the (re-scaled) analytical predictions of \citet[][dashed lines]{Masunaga1998}. The colour coding is the same as in panel (a). (c) Difference between the downstream gas pressure and the upstream ram pressure. (d) Downstream gas pressure (blue), upstream ram pressure (red) and radiation pressures (greens) just upstream (subscript 1) and downstream (subscript 2) of the 
first core accretion shock, in run 69. The upstream radiative fluxes, scaled with the local gas velocity for dimensional coherence, are also plotted for relative comparison in the total 
energy budget (black). (e) First core accretion shock compression ratio for run 69 measured in the 
simulation (black) and predicted for a non-radiative shock (red).}
\label{fig:radius_vs_time}
\end{figure*}

Looking at Fig.~\ref{fig:cores_mass_radius}a, we see that, as expected, all cores are accreting material and growing in mass over time. While the first cores show a 
significant spread in mass at the time of formation (standard deviation of $\sigma = 40$\%), their masses are much more similar at the end of the simulations ($\sigma = 13$\%). In the case of the second cores, we see that all runs show very homogeneous masses in the range $2-4 \times 10^{-3}~\msun$.
The jumps in final first core mass from one run to the next coincide with changes in initial radius $R_{0}$, illustrated by the alternating grey and white background colour bands.
Panel (b) reveals that the simulations produce first cores with very comparable radii. All of them increase 
with time, starting from $\sim$2 AU at core formation, and reaching $\sim$10 AU at the end of the runs. This does contradict some earlier results 
who describe the first core as contracting with time \citep{Masunaga1998,Tomida2013}. To understand this further, we plot the time evolution of the first core radii in 
Fig.~\ref{fig:radius_vs_time}a.

The different colours represent different initial cloud temperatures $T_{0}$ (see legend). We can see that all the runs actually display an initial phase of contraction 
which lasts less than about 10 years, followed by an expansion phase. In panel (b), we plot the first core radii as a function of central density, 
alongside the analytical predictions of \citet{Masunaga1998}. It should be noted that their expressions have been re-scaled to match the runs with $T_{0} = 5$ K; the scaling constant 
of 5.3 AU in their paper has been changed to 2.5 AU\footnote{The quantity $\rho_{\mathrm{ad}}$ has been given the fixed value of $10^{-13}~\text{g~cm}^{-3}$. This 
actually varies between $10^{-13}$ and $10^{-14}~\text{g~cm}^{-3}$ (see Fig.~\ref{fig:all_rhoT}a) which will change the vertical scaling somewhat for different runs, but 
it will not affect the slope.}. The \citeauthor{Masunaga1998} laws actually reproduce the slope of the initial contraction phase very well, and runs with the same initial 
temperature appear to evolve along the same power-law early on. The spread in initial core radius for varying initial cloud temperature goes in the right direction, but 
is larger than the one observed in the simulations.
While the initial contraction phase is in good agreement with previous studies, to understand the subsequent evolution of the first cores further, we isolate run 69, 
which exhibits a behaviour typical of the majority of the models. This run has been highlighted in panels (a) and (b) with a thick black solid curve. 

In Fig.~\ref{fig:radius_vs_time}c, we plot the different pressures on each side of the shock, which enter the momentum 
conservation Rankine-Hugoniot relation for a radiating fluid \citep[see][equation 104.9]{Mihalas1984}; namely the gas internal pressure $P^{\mathrm{gas}}$ (blues), the 
ram pressure $P^{\mathrm{ram}} = \rho u^{2}$ (red/orange), and the radiation pressure $P^{\mathrm{rad}}$ (greens). The subscripts 1 and 2 represent quantities upstream 
and downstream of the shock, respectively. We first notice that the radiation pressure on either 
side of the shock is orders of magitude lower than the other quantities, and does not play a significant role in the momentum and energy exchanges. An accretion shock 
forms when the gas pressure cancels or exceeds the ram pressure of the infalling material. We can see that the post-shock gas pressure $P^{\mathrm{gas}}_{2}$ (bold blue) 
and the pre-shock ram pressure $P^{\mathrm{ram}}_{1}$ (bold red) are comparable at early times ($t < 10$ yr). In fact, $P^{\mathrm{ram}}_{1}$ is slightly higher than 
$P^{\mathrm{gas}}_{2}$, which leads to an early contraction of the first core. For convenience, we have plotted in panel (c) the difference between the downstream gas pressure and the upstream ram pressure (we note from panel d that $P^{\mathrm{gas}}_{1}$ is always at least a factor of 3 below $P^{\mathrm{gas}}_{2}$, and will therefore have virtually no impact on the analysis). Eventually, as the gas pressure inside the core increases, it overcomes $P^{\mathrm{ram}}_{1}$ about 10 years after core formation. This is followed by a phase where $P^{\mathrm{gas}}_{2}$ is noticeably higher than $P^{\mathrm{ram}}_{1}$, causing the core to 
expand. The expansion is halted when the gas and ram pressure become comparable once again, and the radius stabilizes to $\sim6$ AU (small bumps in this and other runs show 
oscillations around hydrostatic equilibrium). While the radiative pressures do increase during this expansion phase, they 
remain much smaller than the hydrodynamic quantities. The radiative pressure is only important in very strong radiative shocks, but the radiative flux, which enters the 
energy conservation Rankine-Hugoniot condition, is important in all radiative shocks \citep[see][section 104 and equation 104.10]{Mihalas1984}. Indeed, we also show for 
relative comparison in Fig.~\ref{fig:radius_vs_time}d the radiative flux (normalised with the local gas velocity for dimensional homogeneity) upstream (black) and 
downstream (grey) of the accretion shock. The radiative fluxes are clearly seen to increase as the shock moves outwards, illustrating the fact that radiative effects become important when the shock reaches larger radii (the spikes in the curves correspond to sign inversions). As the shock moves outwards, the temperature of the 
envelope falls (see the temperature profiles in Fig.~\ref{fig:profiles}c). This means that the dust mean opacity, which scales approximately 
with $T^{2}$, also declines, enabling radiation to escape once the opacity just outside of the accretion shock is small enough.

Another way of looking at the radiative efficiency of the accretion shock is to examine the shock compression ratio, $\mathcal{R}_{\mathrm{sim}} = \rho_{2}/\rho_{1}$. We compare in panel (e) the compression ratio in the simulation (black line) to one predicted from the pre- and post-shock pressures for a purely hydrodynamical gas $\mathcal{R}_{\mathrm{hyd}}$ (red).
Because of radiative effects, we expect $\mathcal{R}_{\mathrm{sim}} \ge \mathcal{R}_{\mathrm{hyd}}$. At early 
times, $\mathcal{R}_{\mathrm{sim}}$ is comparable to its hydrodynamic counterpart. It appears that the shock moves outwards until it 
becomes radiatively efficient, where the compression ratio begins to climb above 5, and the core expansion stalls.
Towards the end of the simulation, a 
second period of expansion is observed. We believe this is due to a combination of two effects. The first comes from a sustained increase in temperature, and hence 
pressure, inside the core (as it continues to accrete material) which implies that $P^{\mathrm{gas}}_{2}$ (bold cyan) rises once again above $P^{\mathrm{ram}}_{1}$ (bold 
red), as seen in panel (c) around a time of 200-300 yr. The evaporation of dust in the centre of the system where temperatures exceed 1500 K could also contribute, as it allows radiation from the hot central region to raise the temperature in the first core. In addition, the first core heats up the envelope (see for instance track 7 in Fig.~\ref{fig:fluid_tracks}a), which leads to a new rise in opacity and optical depth of the envelope, reducing the ability of the accretion shock to radiate accreted energy; 
the fall in compression ratio at late times attests to that fact. 

Returning to Fig.~\ref{fig:cores_mass_radius}, we see that the second core radii are profoundly alike across the board, with an average final size of $4\times 10^{-3}$ AU. We however note that while the first core radii 
increase with time, the second core radii decrease.
\citet{Larson1969}, \citet{Stahler1986} and \citet{Tomida2013}, to name a few, have all reported second Larson cores which grow in size over time. We believe that our 
present results are not inconsistent with these other works, as we have stopped the simulations at very early times, while the second core is still settling. It is very 
possible that the second cores behave like the first cores, with an initial contraction phase, and a subsequent increase in size due to heating and/or mass accumulation 
(which also depletes the accreting envelope).


\subsection{First core formation time and free-fall times}\label{sec:tff_tend}

\begin{figure}
\centering
\includegraphics[width=0.48\textwidth]{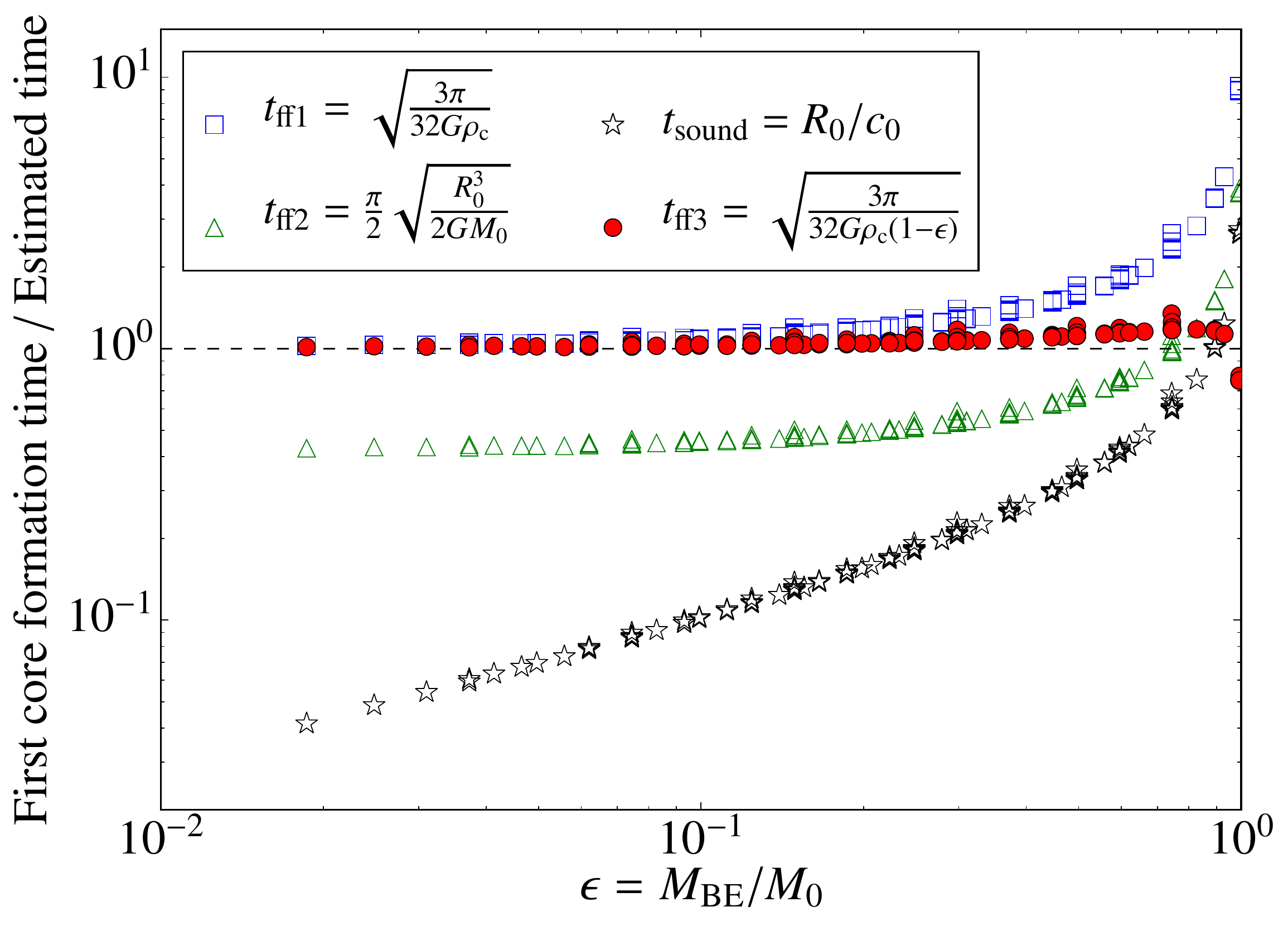}
\caption{Ratio of the time it takes in each run until the first core is formed ($t_{\mathrm{core1}}$) to various collapse time estimates, as a function of $\epsilon$. The blue squares represent the free-fall time computed using the cloud's inital central density $\rho_{\mathrm{c}}$ ($t_{\mathrm{ff1}}$), while the green triangles show the results from an alternate definition of the free-fall time based on the cloud's average density ($t_{\mathrm{ff2}}$). The black stars represent the cloud's initial sound crossing time ($t_{\mathrm{sound}}$). Finally, the best estimate is obtained with a modified version of $t_{\mathrm{ff1}}$ which takes into account how unstable the cloud initially is to gravitational collapse by introducing the $\epsilon$ parameter ($t_{\mathrm{ff3}}$).}
\label{fig:tff_tend}
\end{figure}

In this section, we take a closer look at the collapse time of the cloud, that is the time it takes to form a first Larson core, which we shall call $t_{\mathrm{core1}}$. In comparison, the second core forms almost instantaneously (as we will see in the next sub-section) after the formation of the first core, and the full simulation end time is essentially identical to the first core formation time. We stated in Sect.~\ref{sec:num} that the free-fall time gives a good estimate of the time it takes for the cloud to collapse and form a protostar. We have however found this not to hold in many cases, especially in clouds with a significant amount of thermal support. In Fig.~\ref{fig:tff_tend}, we compare the first core formation times to the free-fall times computed using the cloud's inital central density $\rho_{\mathrm{c}}$ ($t_{\mathrm{ff1}}$, blue squares). While many points lie on the ideal dashed line, there are also a significant number of symbols close to the critical BE-mass at the high-$\epsilon$ end which are at a large distance from it, with up to an order of magnitude in disagreement. An alternate definition of the free fall time, based on the cloud's initial mass and radius ($t_{\mathrm{ff2}}$, green triangles), shows a similar trend, but lies below a ratio of 1 for most of the points, which can be easily understood given the proportionality of the two (cf.~equation~\ref{equ:rhoc_cloud}).
The definition of the free-fall time only considers gravity and ignores pressure forces from the gas, and it is thus not too surprising that in marginally unstable BE spheres, it does not perfectly predict the collapse time.

A timescale which does consider pressure forces is the sound crossing time of the cloud $t_{\mathrm{sound}} = R_{0}/c_{0}$, and this is represented by the black stars in Fig.~\ref{fig:tff_tend}. This timescale also fails to correctly estimate the first core formation time. However, it is obvious that the first core formation time depends on the $\epsilon$ parameter. Indeed, looking at the $t_{\mathrm{ff1}}$ estimates, clouds with low epsilon (i.e.~very unstable initial conditions) lie right on the dashed line, when gravity is the dominant force. When clouds have an $\epsilon$ close to unity, where pressure support and gravity forces are comparable, it takes longer for the collapse to initiate. We found that correcting the $t_{\mathrm{ff1}}$ to now read
\begin{equation}\label{equ:tff3}
t_{\mathrm{ff3}} = \sqrt{\frac{3\pi}{32 G \rho_{\mathrm{c}}(1-\epsilon)}}
\end{equation}
gives a very good estimate of the first core formation time; this is represented by the red circles in Fig.~\ref{fig:tff_tend}. The free-fall timescale is dominated by the first stages of the contraction, as the collapse time becomes shorter and shorter the higher the density gets. It therefore makes sense that, at early times, the effective gravitational pull is given by that which is unbalanced by the pressure gradient. One would then expect the correct density to enter the formula for the free-fall time to be the differential density compared to the critical BE density: $\rho_{\mathrm{c}} - \rho_{\mathrm{c,crit}} = \rho_{\mathrm{c}} - \rho_{\mathrm{c}} \epsilon$, which is exactly what we find in equation~(\ref{equ:tff3}).
We can also see that as $\epsilon$ goes to 0, we recover the original definition $t_{\mathrm{ff1}}$, while as $\epsilon$ reaches unity, $t_{\mathrm{ff3}}$ goes to infinity and the cloud does not collapse.


\subsection{First core lifetimes}\label{sec:first_core_lifetimes}

\begin{figure*}
\centering
\includegraphics[width=\textwidth]{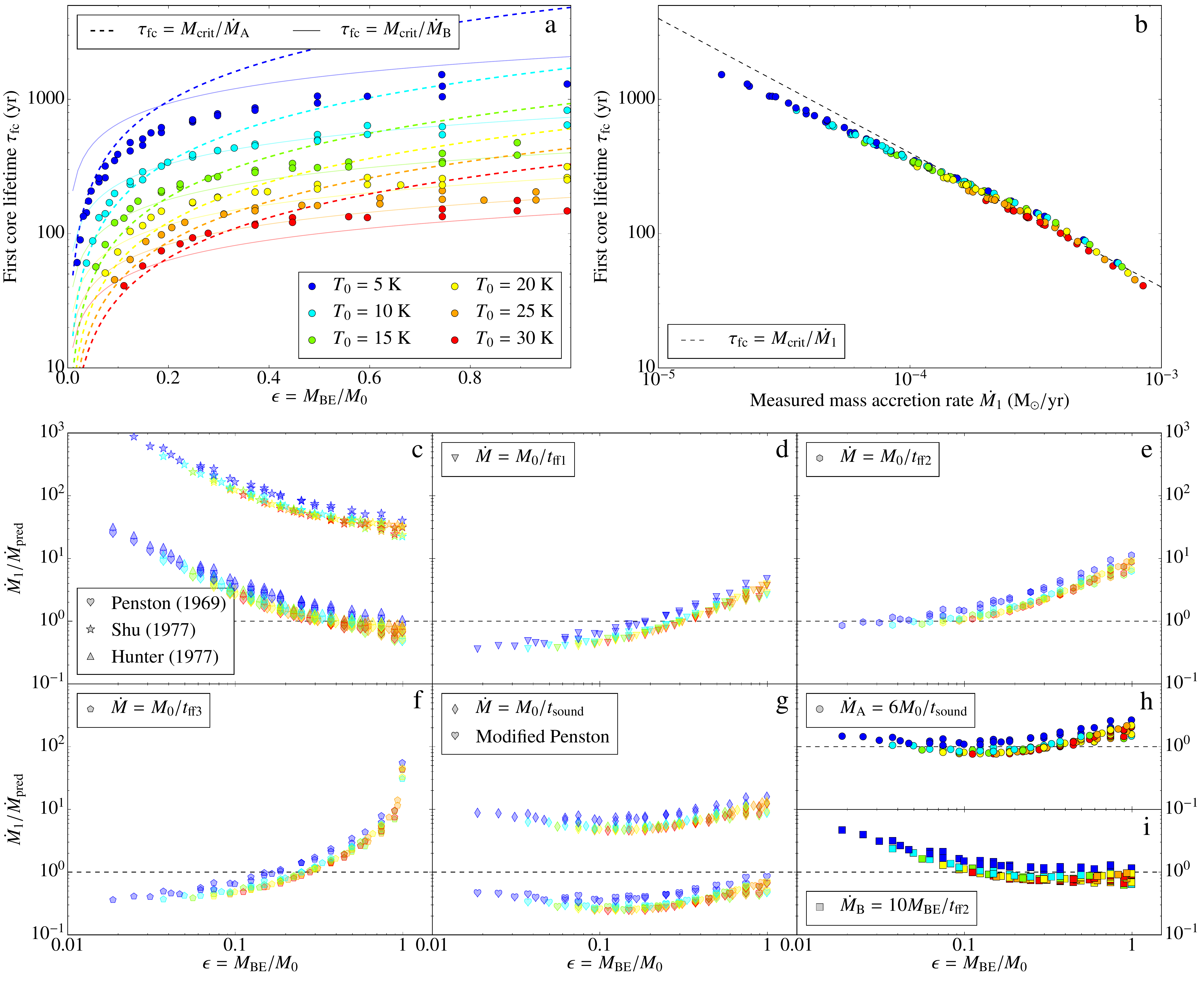}
\caption{(a) First core lifetime $\tau_{\text{fc}}$ as a function of $\epsilon$ measured in the simulations (circles).  The dashed and thin solid lines represent our analytical predictions (see text). (b) First core lifetime as a function of mass accretion rate for all runs. The black dashed 
line portrays a simple estimate from dimensional arguments (see text). (c)-(i) Ratio of the mass accretion rate measured in the simulation $\dot{M_{1}}$ to various 
predictions $\dot{M}_{\mathrm{pred}}$ from numerical and similarity solutions, as a function of $\epsilon$. (c) The hearts represent the similarity solution of \citet{Larson1969} and \citet{Penston1969}, the stars 
are the solution of \citet{Shu1977}, while the triangles are from the results of \citet{Hunter1977}. In panels (d) to (f), the $\dot{M}_{\mathrm{pred}}$ estimate is obtained from dividing the initial cloud mass $M_{0}$ by the free-fall times $t_{\mathrm{ff1}}$, $t_{\mathrm{ff2}}$, and $t_{\mathrm{ff3}}$. (g) The hearts represent a modified version of \citeauthor{Penston1969}'s formula, which is equivalent, barring a scaling factor, to the prediction using the cloud sound crossing time (diamonds). (h) The circles represent the first best fitting formula $\dot{M}_{\text{A}}$ we have used to predict the lifetimes. (i) The squares show the second mass accretion rate estimate $\dot{M}_{\text{B}}$ we have selected for our analysis. The colours in all panels code for different initial 
temperatures (see legend in panel a).}
\label{fig:lifetimes}
\end{figure*}

This section focuses on the time elapsed between the formation of the first and second cores, which we will refer to as the `first core lifetime'. Once the second core
is formed, the observable properties of the system such as spectral energy distributions dramatically change, due to the high temperatures inside the second core 
\citep[see for example][]{Masunaga2000}\footnote{As \citet{Masunaga2000} show, it may take some time before the observable properties of the protostellar 
system, such as the spectral energy distributions, are altered. They predict that this does not happen immediately after the second core is formed, but well into the main 
accretion phase.}. The first core lifetimes for all the runs
are displayed in Fig.~\ref{fig:lifetimes}. The top panel (a) presents the lifetimes as a function of the initial mass to Bonnor-Ebert mass ratio $\epsilon$. For $\epsilon \gtrsim 0.4$, the lifetimes appear to depend only weakly on $\epsilon$, showing a spread between 100 and 1000 years. Below, the lifetimes drop with $\epsilon$. 

If one assumes that the first core is in quasi-hydrostatic equilibrium, one can see that, if it is continuously accreting material from its outer envelope, its 
temperature will also steadily increase. It is also safe to assume that the higher the mass accretion rate, the faster it will heat up and enter the second phase of 
collapse. This therefore predicts a correlation between first core lifetime and mass accretion rate, and we show this in Fig.~\ref{fig:lifetimes}b, where the tight 
correlation is glaring (the value for the accretion rate is taken as the average rate over the lifetime of the first core).
Assuming a uniform density and temperature, the equation of hydrostatic equilibrium for a sphere is
\begin{equation}
\frac{dP}{dr} = -\rho G \frac{M}{r^{2}} ~.
\end{equation}
Approximating $dP/dr = (P_{0} - P_{\mathrm{c}})/R$ where $P_{\mathrm{c}}$ is the central core pressure, $P_{0}$ ($\ll P_{\mathrm{c}}$) is the outside pressure, and $R$ the core radius, we find
\begin{equation}
P_{\text{c}} = \rho G \frac{M}{R} ~,
\end{equation}
which we finally convert to temperature to obtain
\begin{equation}\label{equ:Tc}
T_{\text{c}} =  \frac{\mu m_{\text{H}}}{k_{\mathrm{B}}} \frac{GM}{R} ~.
\end{equation}
If we assume the radius of the first core remains approximately constant, there is a critical mass $M_{\text{crit}}$ that the first core must reach, so that $T_{\text{c}} = 2000$ K. This is further justified by the fact that all first cores appear to reach very similar masses at the end of the simulations in Fig.~\ref{fig:cores_mass_radius}a. In fact, by setting $R = 5$ AU in equation~(\ref{equ:Tc}), we obtain $M_{\text{crit}} = 0.04~\msun$, which is fully consistent with Fig.~\ref{fig:cores_mass_radius}a. We can now define the first core lifetime as being the time it takes for the core to reach $M_{\text{crit}}$, having started at an 
initial mass $M_{\text{init}}$ at the time of formation. One can go even further by neglecting $M_{\text{init}}$ altogether as it is on average 10 times lower than $M_{\text{crit}}$, allowing us to write the very simple relation
\begin{equation}\label{equ:taufc}
\tau_{\text{fc}} =  \frac{M_{\text{crit}}}{\dot{M_{1}}} ~,
\end{equation}
where $\dot{M_{1}}$ is the first core mass accretion rate. This relation (for $M_{\text{crit}} = 0.04~\msun$) is plotted with a dashed black line in Fig.~\ref{fig:lifetimes}b, alongside the measured first 
core lifetimes and mass accretion rates from the simulations. Although not perfect, the law is a decent fit to the measured lifetimes, considering the simplicity of the 
argument we have used.

To try and estimate the first core lifetimes from the initial conditions, we now only need to predict the first core mass accretion rates, as they are fully correlated to 
the lifetimes. Many authors have studied numerically and analytically the problem of an isothermal sphere of gas, collapsing under its own gravity. As discussed in 
\citet{Foster1993}, several estimates of mass accretion rates were derived by \citet{Larson1969}, \citet{Penston1969}, \citet{Shu1977} and \citet{Hunter1977}. They are 
expressed as a function of the cloud's isothermal sound speed $\dot{M}_{\text{pred}} = K c_{0}^{3} / G$, where $K$ is a constant equal to 47 in the 
\citeauthor{Larson1969}-\citeauthor{Penston1969} solution, 0.975 for \citeauthor{Shu1977}, and 36 for \citeauthor{Hunter1977}. These values were computed at the epoch of first core formation, and are not constant in time \citep[see][for more details]{Hunter1977}, but are useful to obtain order of magnitude estimates. We show these predictions in Fig.~\ref{fig:lifetimes}c, where the mass accretion rates $\dot{M_{1}}$ measured in the simulations are compared to the aforementioned predictions, as a function of $\epsilon$. All three distributions have the same shape, and show a distinct departure from the measured mass accretion rate at low $\epsilon$. 
We also note here that \citeauthor{Shu1977}'s prediction performs much worse that the other two, with ratios always higher than 10. A different 
way of looking at this problem is to estimate the mass accretion rate through a very simple dimensional analysis. We suppose that the sphere of gas collapses onto itself in 
a free-falling manner, meaning that all the mass will reach the centre of the system in one free-fall time. The mass accretion rate is then simply the mass of the cloud 
divided by the free-fall time
\begin{equation}\label{equ:mdot_pred}
\dot{M}_{\text{pred}} = \frac{M_{0}}{t_{\text{ff}}} ~.
\end{equation}
The predictions for the three different definitions of the free-fall time from Sect.~\ref{sec:tff_tend} are shown in Figs~\ref{fig:lifetimes}d-f. While the dimensional estimate works better than the predictions in panel (c), we can see that a significant amount of spread above and below a ratio of unity is still present, especially in the $t_{\mathrm{ff3}}$ case, showing that even a good estimate of the cloud collapse time does not provide a reliable measure of the mass accretion rate. The final timescale we used in Sect.~\ref{sec:tff_tend} was the sound crossing time, and the mass accretion rate obtained from $\dot{M}_{\text{sound}} = M_{0}/t_{\text{sound}}$ is shown in Fig.~\ref{fig:lifetimes}g (diamonds). This time, the distribution is close to horizontal; it only lacks a scaling constant to bring it onto the dashed line. In fact, a similar expression can be found from the \citeauthor{Penston1969} solution, if we take into account the effect of the $\epsilon$ parameter. Indeed, the increasing ratio with decreasing $\epsilon$ panel (c) prompted us to attempt to correct the formulas by incorporating $\epsilon$ in the expressions. Simply dividing the \citeauthor{Penston1969} expression by $\epsilon$ yields
\begin{equation}\label{equ:mdot_penston_mod}
\dot{M}_{\text{pred}} = 47 \frac{c_{0}^{3}}{G\epsilon} = \frac{47}{2.4} M_{0} \frac{c_{0}}{R_{0}} ~,
\end{equation}
which is identical to $\dot{M}_{\text{sound}}$, except for a factor of $\sim$20. This expression also implies that for an infinite cloud mass with a given radius, $\epsilon \rightarrow 0$ and the mass accretion rate goes to infinity. We have ultimately decided to use a scaled version of $\dot{M}_{\text{sound}}$ (which is the same as scaling the modified Penston estimate) as it showed the best agreement with the measured mass accretion rates, with a factor of 6,
\begin{equation}\label{equ:mdot_A}
\dot{M}_{\text{A}} = 6 M_{0} / t_{\text{sound}} ~,
\end{equation}
as shown in Fig.~\ref{fig:lifetimes}h.
This constant probably accounts for the fact that the mass accretion rate onto the core is not constant in time, but begins slowly, and accelerates as the density increases in a 
runaway fashion during the early stages of the collapse. The value of 6, however, implies our dimensional analysis agrees within an order of magnitude with the numerical 
models. The prediction is solid for $\epsilon \leq 0.3$, above which it starts to depart from the ideal dashed line.
A second attempt at accurately predicting the first core mass accretion rates was made by correcting the expressions in panels (d)-(f) with the $\epsilon$ parameter. Multiplying the initial mass $M_{0}$ by $\epsilon$ is equivalent to using the BE mass in the numerator of the $\dot{M}_{\text{pred}}$ formula, and again using a scaling constant, we found
\begin{equation}\label{equ:mdot_B}
\dot{M}_{\text{B}} = 10 M_{\text{BE}} / t_{\text{ff2}}
\end{equation}
to give good results, as shown in Fig.~\ref{fig:lifetimes}i. This time the agreement with the measured mass accretion rates is excellent at high $\epsilon$, and only starts to break down for $\epsilon < 0.2$.

Returning to the top panel of Fig.~\ref{fig:lifetimes}, we can now express the mass accretion rate, and hence the first core lifetime, in terms of $\epsilon$.
This yields
\begin{flalign}
\tau_{\text{fc}}^{\text{A}} = \, & \displaystyle M_{\text{crit}}~\frac{1}{6}~\frac{1}{c_{0}}~\frac{R_{0}}{M_{0}} \label{equ:tau_ana1}\\
                            = \, & \displaystyle M_{\text{crit}}~\frac{1}{6}~\frac{1}{c_{0}}~\frac{G \epsilon}{2.4 c_{0}^{2}} \label{equ:tau_ana2}\\
                            \simeq \, & \displaystyle 1700~ \left(\frac{M_{\text{crit}}}{0.04~\msun}\right) \left(\frac{\mu}{2.31}\right)^{3/2}  \epsilon \left(\frac{T_{0}}{10~\text{K}}\right)^{-3/2} ~\text{yr} ~, \label{equ:tau_ana3}
\end{flalign}
when using the $\dot{M}_{\text{A}}$ estimate in equation~\ref{equ:mdot_A}. This relation is represented in Fig.~\ref{fig:lifetimes}a by the dashed coloured lines for different values of 
$T_{0}$. The temperature spread in lifetimes for a given $\epsilon$ is well reproduced by the analytical function. At the high-$\epsilon$ end, the analytical description over-estimates the lifetimes, as the $\dot{M}_{\text{A}}$ prediction worsens for the most stable runs (as seen in the right hand side of panel h). Using the $\dot{M}_{\text{B}}$ expression in equation~\ref{equ:mdot_B} gives
\begin{flalign}
\tau_{\text{fc}}^{\text{B}} = \, & \displaystyle M_{\text{crit}}~\frac{\pi}{\sqrt{8G}}~\frac{1}{10\epsilon}~\left(\frac{R_{0}}{M_{0}}\right)^{3/2} \label{equ:tau_ana4}\\
                            = \, & \displaystyle M_{\text{crit}}~\frac{\pi G}{10\sqrt{8}}~\left(2.4 c_{0}^{2}\right)^{-3/2} \epsilon^{1/2} \label{equ:tau_ana5}\\
                            \simeq \, & \displaystyle 740~ \left(\frac{M_{\text{crit}}}{0.04~\msun}\right) \left(\frac{\mu}{2.31}\right)^{3/2}  \epsilon^{1/2} \left(\frac{T_{0}}{10~\text{K}}\right)^{-3/2} ~\text{yr} ~, \label{equ:tau_ana6}
\end{flalign}
which is plotted in Fig.~\ref{fig:lifetimes}a with thin solid lines. Once again the range of temperatures is adequately covered, but this time the fit is much better towards high values of $\epsilon$, where the $\epsilon^{1/2}$ slope follows the data much more closely. The two relations are equal for $\epsilon_{\mathrm{AB}} = 9\pi^{2}/(25 \times 8 \times 2.4) = 0.185$, and they are a good match to the data in their respective $\epsilon$ ranges. 


\subsection{Further correlations}\label{sec:correlations}

To extract further correlations between the different simulation initial parameters and final core properties from Table~\ref{tab:parameters}, we computed for each column 
in the table the correlation of the data in that column to all the other columns. The correlation between columns $x$ and $y$ is defined by
\begin{equation}
r_{xy} = \frac{\mathrm{cov}(x,y;w)}{\sqrt{\mathrm{cov}(x,x;w)~\mathrm{cov}(y,y;w)}} ~,
\end{equation}
where the covariance is given by
\begin{equation}
\mathrm{cov}(x,y;w) = \frac{\sum_{i} w_{i} (x_{i}-\mathrm{m}(x;w)) (y_{i}-\mathrm{m}(y;w))}{\sum_{i}w_{i}} ~,
\end{equation}
and
\begin{equation}
\mathrm{m}(x;w) = \frac{\sum_{i} w_{i} x_{i}}{\sum_{i}w_{i}}
\end{equation}
is the weighted mean. The subscript $i$ denotes a particular run. The weights are used to minimise the influence of outlier data points, and the method used to compute the weights is given in detail in Appendix~\ref{app:corr_weighting}.
This correlation matrix $r_{xy}$ is plotted in Fig.~\ref{fig:correlations} (it should be noted that the arrays $x$ and $y$ actually represent the logarithm 
of the quantities in Table~\ref{tab:parameters}). Within the initial parameters, some obvious strong correlations appear, such as $(\rho_{\text{c}},t_{\text{ff1-2}})$ or $(\rho_{\text{c}},R_{0})$. It is also self-evident that the simulation end time (or $t_{\text{ff3}}$) should correlate with the other free-fall times. Some correlations also emerge between 
the run number and various parameters related to the initial radius, which comes from the manner in which we have ordered the runs from lower to higher radii.

\input{correlations.tex}

Among the first and second core parameters, we also unsurprisingly find that accretion luminosity is correlated to the mass accretion rate, since the accretion luminosity 
is computed from the mass accretion rate. It also relates to the mass and radius of the cores.
As in Sect.~\ref{sec:first_core_lifetimes}, we find once again that the first core lifetime $\tau_{\text{fc}}$ is strongly correlated to the mass accretion rate $\dot{M_{1}}$. More interestingly, it appears to also be correlated to the first core radius. As we have seen in Sect.~\ref{sec:first_core_lifetimes}, the first core grows in size 
with time for most runs, and it seems obvious that the longer the first core lives, the bigger it will get. This correlation is actually visible in the top right corner of 
Fig.~\ref{fig:radius_vs_time}a, where the final radius $R_{1}$ increases for longer lived cores. The luminosity at the outer radius of the simulated system $L_{\mathrm{out}}$ appears quite solidly related to, among others, the first core mass, as result fully expected from fundamental stellar physics where stars are considered to be radiating spheres of gas \citep[see for example][chapter 1]{Phillips1999}. Fitting a power law to the luminosities yielded the relation $L_{\mathrm{out}}\propto M_{1}^{3.25}$, which is close to the value of 3.5 commonly quoted in text books.

In the second core properties, the core radius naturally correlates well to the accretion and radiative luminosities ($r > 0.85$) as it enters the calculation of both ($L^{\text{acc}} = G M \dot{M}/R$ and $L^{\mathrm{rad}} = \pi R^{2} F$), while the core mass has relatively strong relationships with all of the second core properties ($r > 0.7$). The accretion luminosity and the radiative luminosity are also correlated, even though they differ by many orders of magnitude; they are most probably related via the core radius. Strangely enough, the $\epsilon$ parameter does not seem to strongly correlate, on its own, with the first and second core properties. As it was found to be an important parameter in the preceding sections, this illustrates that using correlations alone are not enough to extract a complete physical understanding of the collapse of spherical clouds.

There are many possible relations to be interpreted from this data, and going through them all would be too lengthy for the present article. Each square in the figure (coloured or grey) links to a figure on the online server which displays the $x$ quantity plotted against the $y$, providing the reader with the entire collection of plots.


\section{The collapse of a uniform density cloud}\label{sec:uniform_density}

Throughout past studies of the collapse of an isolated sphere of gas, two main types of initial conditions have been used. The first, as was employed in this work, is a super-critical Bonnor-Ebert sphere. The second is simply a sphere with a uniform density. There are many arguments against how a uniform density sphere could form in nature, but a such a body was found to very rapidly adopt a Bonnor-Ebert like profile once it started to collapse `forgetting' its initial state \citep[see for instance][]{Larson1969,Masunaga1998}, and the subsequent evolution was deemed not to depend on the initial density profile. Some authors, including \citet{Masunaga2000}, however argued that the density profile does affect the protostellar evolution, mass accretion rates and luminosities, due to different dynamics. In this section, we re-run some of our simulations by adopting a uniform density as a starting point, and compare them to our standard runs.

\begin{figure}
\centering
\includegraphics[width=0.48\textwidth]{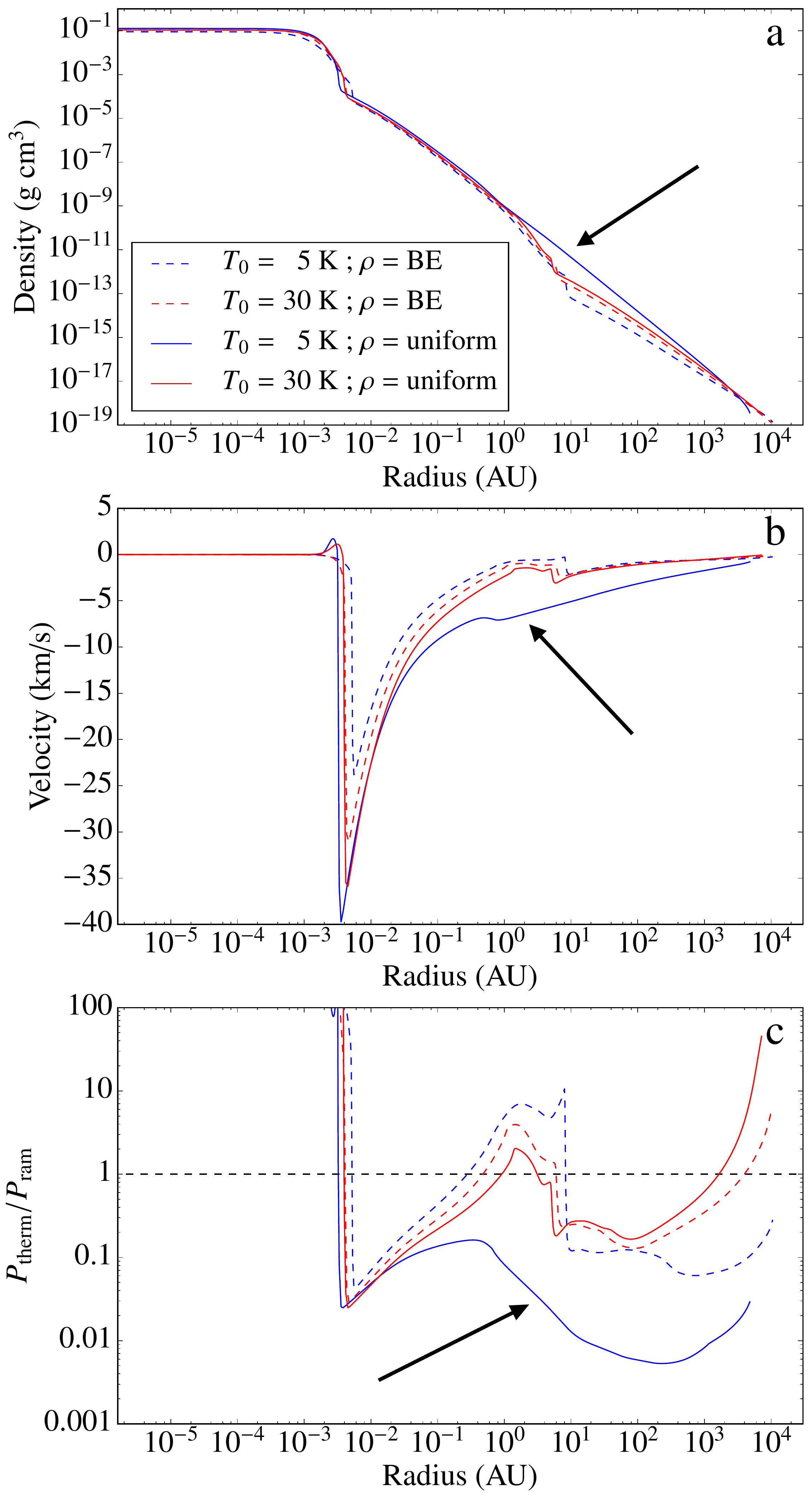}
\caption{(a) Final density profiles for Bonnor-Ebert (dashed) and uniform density (solid) set-ups with initial temperatures of 5 K (blue) and 30 K (red). (b) Same as in (a) but for the gas velocity. (c) Ratio of the gas thermal pressure to ram pressure as a function of radius. The arrows point to the absence of a first Larson core in the most unstable uniform density run.}
\label{fig:uniform_density}
\end{figure}

We selected intermediate runs 95 to 100, which all have an initial mass of $4~\msun$, initial radius of 12,000 AU, and varying temperatures.
We show the results of the uniform density simulations in Fig.~\ref{fig:uniform_density} (solid lines), alongside the original BE runs (dashed), for initial temperatures of 5 and 30 K. We first note that the initial density profile does not significantly affect the final density and velocity profiles, as well as the sizes and masses of the protostellar cores, when $T_{0} = 30$ K. This was found to hold true for all initial temperatures above 5 K. The case of $T_{0} = 5$ K is however intriguing, as it appears to be devoid of a first core accretion shock. We observe an accretion shock at the second core border, but the flow is continuous throughout the rest of the infalling envelope.
There are no discontinuities in either density or velocity, as indicated by the arrows in Fig.~\ref{fig:uniform_density}. The system has proceeded directly to 
the second collapse, before forming a hydrostatic body during the first adiabatic phase. This occurs when the gas at the centre of the system rapidly reaches the dissociation temperature of $\text{H}_{2}$ molecules $T_{\text{H}_{2}} = 2000$ K, due to a strong compressional heating. It can also be understood in terms of when a cloud is initially highly unstable (because it has very little initial thermal support with such a low initial temperature of 5 K), the strong ram pressure due to the high 
infalling velocity is always above the thermal pressure inside the core, and an accretion shock never forms. This is illustrated in Fig.~\ref{fig:uniform_density}c which 
displays the ratio of thermal pressure $P_{\text{them}}$ to the ram pressure $P_{\text{ram}} = \rho u^{2}$ as a function of radius at the end of the simulations. For the uniform $T_{0} = 5$ K run, unlike the other three runs, the thermal pressure never counter-balances the ram pressure from the infalling gas in the region where the first core accretion shock typically forms ($\sim$1--10 AU). Contrastingly, all runs have a clear accretion shock that forms at the second Larson core border ($3-5\times 10^{-3}$ AU). 

After having run many more models with a uniform density profile ($\sim$90 additional runs), only the most unstable ($\epsilon < 0.1$) set-ups were found to lack a first Larson core. The other runs behaved very similarly to the BE models. This does not invalidate previous studies which were using a uniform initial density, as the clouds were never unstable enough to skip the first core stage. We believe that this peculiar behaviour is because with an initial uniform density, the free-fall time is the same everywhere in the cloud. This implies that all the mass should reach the centre at the same time. The free-fall time considers gravity to be the only important force, ignoring gas pressure forces, which are in fact still able form a first core, even when the free-fall time is homogeneous. On the other hand, when gravity strongly dominates in very unstable clouds, the strong infalling motions do not permit the formation of a first hydrostatic core. The pressure inside the second Larson core is so high that the collapse is eventually halted, but the system's density and velocity profiles strongly resemble that of a singular isothermal sphere.


\section{Comparison with observations}\label{sec:observations}

\begin{figure}
\centering
\includegraphics[width=0.48\textwidth]{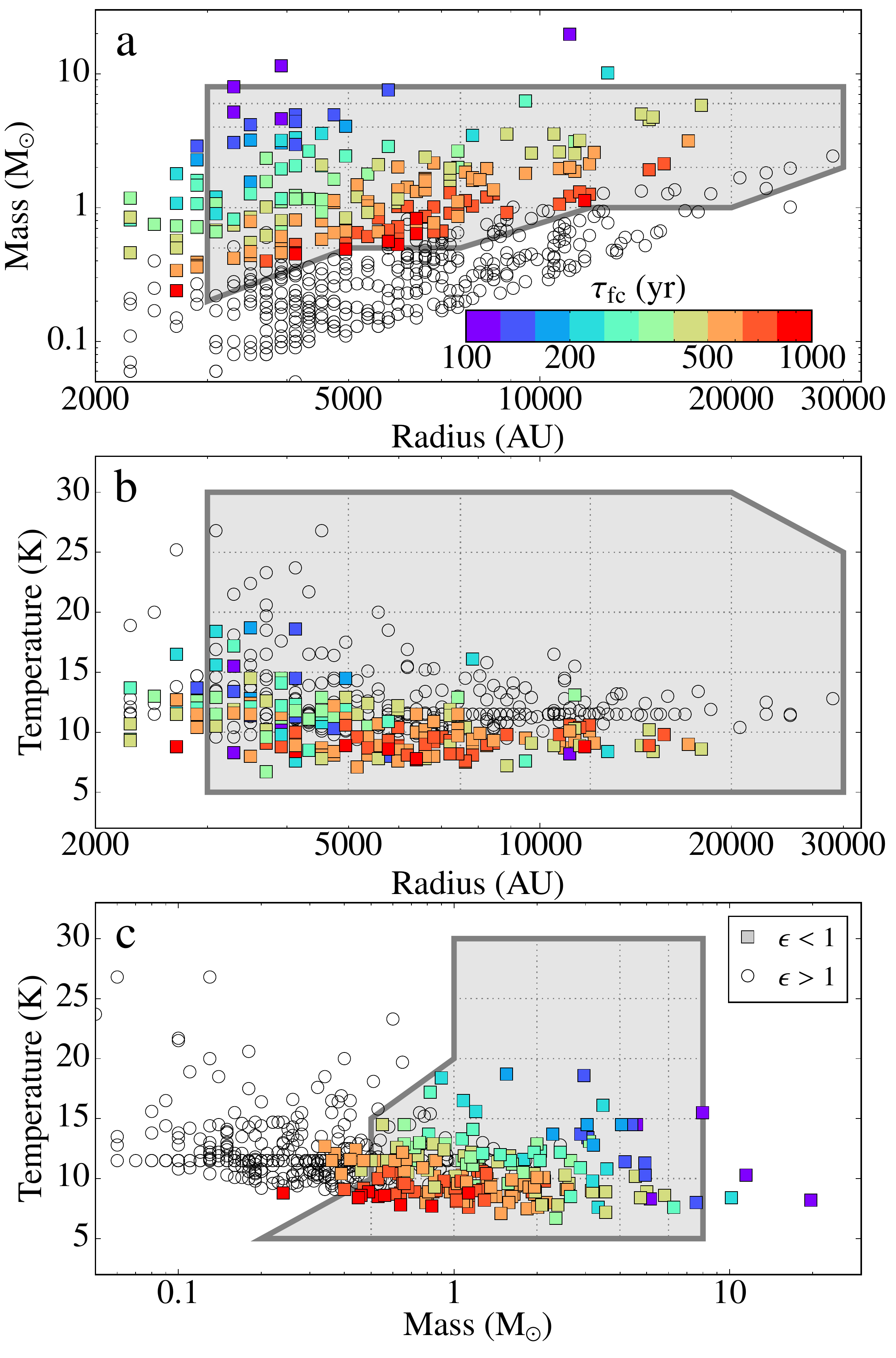}
\caption{Comparison between the radii, temperatures and masses of the prestellar cores in the Aquila molecular cloud and the parameter space covered by the initial 
conditions of our set of numerical simulations. In each panel, the empty circles represent observed cores with $\epsilon > 1$, while the coloured squares are for cores 
with $\epsilon < 1$. The colour of the squares code for their predicted first core lifetimes. The grey area deliniates the simulation parameter space: every corner or 
intersection of a dotted line with either another dotted line or the thick grey boundary represents one simulation.}
\label{fig:parameter_space}
\end{figure}

The initial conditions of our simulations can be compared to the observations of \citet{Konyves2015} who performed an extensive survey of dense globules\footnote{These 
star-less density fluctuations in molecular clouds are often called `dense cores', which represent an evolutionary stage analogous to the initial conditions in our set-up. 
We shall call them `globules' here, to make a distinction between these prestellar cloud cores and the first and second Larson cores, which are much smaller and denser 
objects.} in the Aquila cloud complex. We selected from their sample all the globules that were marked as `prestellar' and `protostellar', and plotted them in Fig.~\ref{fig:parameter_space}, where we have made use of the deconvolved radii in their database. In each panel, the squares and circles represent the 
observations, while the grey area deliniates the region in the parameter space covered by our simulations.
The empty circles symbolise observed globules with $\epsilon > 1$, which we predict will not be collapsing under the influence of gravity alone. Their masses are below 
the Bonnor-Ebert critical value, and they will only form protostars if the external (thermal or kinetic) pressure increases their density, or if they are allowed to cool down 
until the internal pressure gives in to gravity. Most empty circles actually lie outside of our parameter space coverage, as we have, for obvious reasons, decided to only 
simulate systems with a mass higher than $M_{\text{BE}}$.

The colours of the squares represent the estimated lifetimes of their first cores before the second collapse takes place, obtained from our analytical estimates in equations (\ref{equ:tau_ana3}) and (\ref{equ:tau_ana6}), where we have used $\tau_{\text{fc}}^{\text{A}}$ for $\epsilon < \epsilon_{\text{AB}}$ and $\tau_{\text{fc}}^{\text{B}}$ elsewhere. As all the coloured globules have 
temperatures below 20 K, we estimate that almost all first cores will live for at least 100-200 years, but only 50\% have a predicted lifetime above 500 years. With no temperatures below 6 K, no first core will live longer than 1000 years. We note that our predicted first core lifetimes are lower limits, as rotational and magnetic support in 3D set-ups will stretch the lifetimes by reducing the mass accretion rates.

There is a certain number of coloured globules which lie outside of the grey area, at the low-radius (high-density)
end of the dataset. We did not extend our parameter space too far in that direction, as we deemed initial densities above $\sim$10$^{-16}~\text{g~cm}^{-3}$ to be too high 
for realistic initial conditions. In hindsight, it appears that such high densities are observed in molecular clouds, but are probably the result of strong compression 
from interstellar turbulence, shocks, or colliding flows. Exploring the high-mass high-density regime with our simulations may not be the most interesting, or realistic, 
as turbulence most probably induces low levels of fragmentation of the globule in high-mass star formation \citep{Commercon2011,Palau2014}, which cannot be captured with 
a 1D model. However, the very low-mass end of the high-density region may represent a pathway to forming brown dwarfs. It is always difficult to form brown dwarfs in 
spherically symmetric simulations, as the globules must have either a very low temperature, or a very high density (or both) to be able to collapse with such a small 
mass\footnote{This is precisely the origin of the low-mass turn-over in the initial mass function \citep{PadoanNordlund2002}}. Indeed, coloured globules with masses of the order of 0.1~\msun and radii below 3000 AU are absent from the observations. It is in fact believed 
that one actually starts with a higher mass globule which is more likely to exceed $M_{\text{BE}}$, and a significant portion of the initial mass is subsequently ejected 
through a powerful outflow, as suggested by the simulations of \citet{Machida2009}. We have thus decided not to extend our parameter space to radii below 3000 AU.

We believe that our simulation set adequately covers the conditions typically found in molecular clouds (and that remain reasonable to simulate with a spherically 
symmetric scheme), and we have even explored temperatures higher than most of the observed globules in the Aquila cloud. Our higher temperature models will be relevant for e.g.~inner Galaxy and outer Galaxy clouds that are more irradiated due to nearby massive stars \citep{Lindberg2012}. The relations found in this study can be used to 
make evolutionary predictions on observed prestellar systems, such as the time they will take to form a first core, and the lifetimes of these cores. The typical lifetime of the first core will also give an indication of the probability of observing a first core object in a given field. Our analytical predictions suggest that one has a better chance of observing a first Larson core in low-temperature systems, where the time span between the first and second core formation is the longest.


\section{Conclusions}

An extensive set of simulations modelling the gravitational collapse of a dense cloud has been carried out. The calculations follow the initial isothermal contraction of 
the sphere, the transition into the first adiabatic phase, the dissociation of $\text{H}_{2}$ molecules, the second collapse, and finally the early stages of the second adiabatic phase. They can also be used to trace the evolution of individual pieces of fluid, as they fall inwards, travel through accretion shocks, and/or get heated via radiation from the central object.
The 143 runs show a 
wide spread in the thermal evolutionaly tracks ($\rho_{\mathrm{c}}$ vs $T_{\mathrm{c}}$) they follow; this is due to differing initial temperatures, and also to 
different optical depths which regulate the radiative cooling. The models also revealed that while some dispersion was observed in the density, velocity, or temperature radial profiles outside of the first core accretion shock, the profiles showed very little scatter inside hydrostatic body where they seemed to `forget' their initial conditions.

Considering the wide ranges of masses and densities covered by the parameter space, the simulations produced first and second cores with remarkably similar properties. 
The average first core mass and radius are $\overbar{M_{1}} = 4 \times 10^{-2}~\msun$ and $\overbar{R_{1}} = 7$ AU, while the average second core mass and radius are $\overbar{M_{2}} = 3 \times 10^{-3}~\msun$ and $\overbar{R_{2}} = 4 \times 10^{-3}$ AU. Within less than an order of magnitude, these quantities were basically independent of the initial conditions. In our simulations, the first cores were observed to contract for a brief period before expanding until the accretion shock at their border became radiatively efficient, where the (mostly kinetic) energy of the infalling material impacting onto the core surface could be radiated away.

A curious behaviour was observed when adopting a uniform initial density profile instead of the BE solution; the most unstable runs were found to be devoid of a first Larson core. After a careful study of this phenomenon, we concluded that it was an artefact of the initial conditions where the free-fall time is the same in all parts of the cloud. Along with the fact that it seems unlikely for a uniform density body to form in nature, we believe the BE simulations are a better representation of the evolution of collapsing dense clouds.

A comprehensive study of the lifetimes of the first Larson cores (i.e.~the time elapsed between the formation of the first and second cores) showed that they strongly depended 
on the first core mass accretion rate. Two approximate predictive laws linking the initial conditions to the first 
core lifetimes were also derived.
We note that the lifetimes given here are lower-bounds on realistic estimates as magnetic and kinetic support 
(rotational or turbulent) in 3D situations will slow down the core evolution, reducing the mass accretion rates. The flow of mass will also proceed via the formation of 
accretion discs, as opposed to covering the entire $4\pi$ steradian surface of the core. It is, however, interesting to mention that \citet{Tomida2013} found first 
core lifetimes to range between 700 and 1000 years ($\epsilon = 0.8$), while \citet{BateTricco2014} observed 50-100 years ($\epsilon = 0.14$) in their simulations, which 
is not too different from our 1D results, even though both works consider a 3D magnetised cloud with rotation.
Magneto-hydrodynamic simulations (ideal or with Ohmic dissipation) of rotating clouds probably result in similar first core lifetime because of the efficient angular momentum transport provided by magnetic braking. Without magnetic fields, the lifetime can be considerably longer. Recent non-ideal MHD simulations including ambipolar diffusion \citep{Tomida2015} show longer lifetimes, as the resistive effects suppress angular momentum transport.



\begin{acknowledgements}
NV gratefully acknowledges support from the European Commission through the Horizon 2020 Marie Sk{\l}odowska-Curie Actions Individual Fellowship 2014 programme
(Grant Agreement no. 659706).
TH is supported by a Sapere Aude Starting Grant from the Danish Council for Independent Research.
We thank B.~Commer\c{c}on and J. Masson for very useful discussions during the writing of this paper.
We would also like to express our gratitude to the referee, whose extremely constructive comments were much appreciated and have greatly improved this manuscript.
This work made use of the astrophysics HPC facility at the University of Copenhagen, which is supported by a research grant (VKR023406) from VILLUM FONDEN.
We finally thank the 
Service d'Astrophysique, IRFU, CEA Saclay, and the Laboratoire Astrophysique Interactions Multi-\'{e}chelles, France, for granting us access to the
supercomputer \texttt{IRFUCOAST} where some of the calculations were also performed.
\end{acknowledgements}


\bibliographystyle{aa}


\appendix

\section{Resolution study}\label{app:resolution}

\begin{figure*}
\centering
\includegraphics[width=0.9\textwidth]{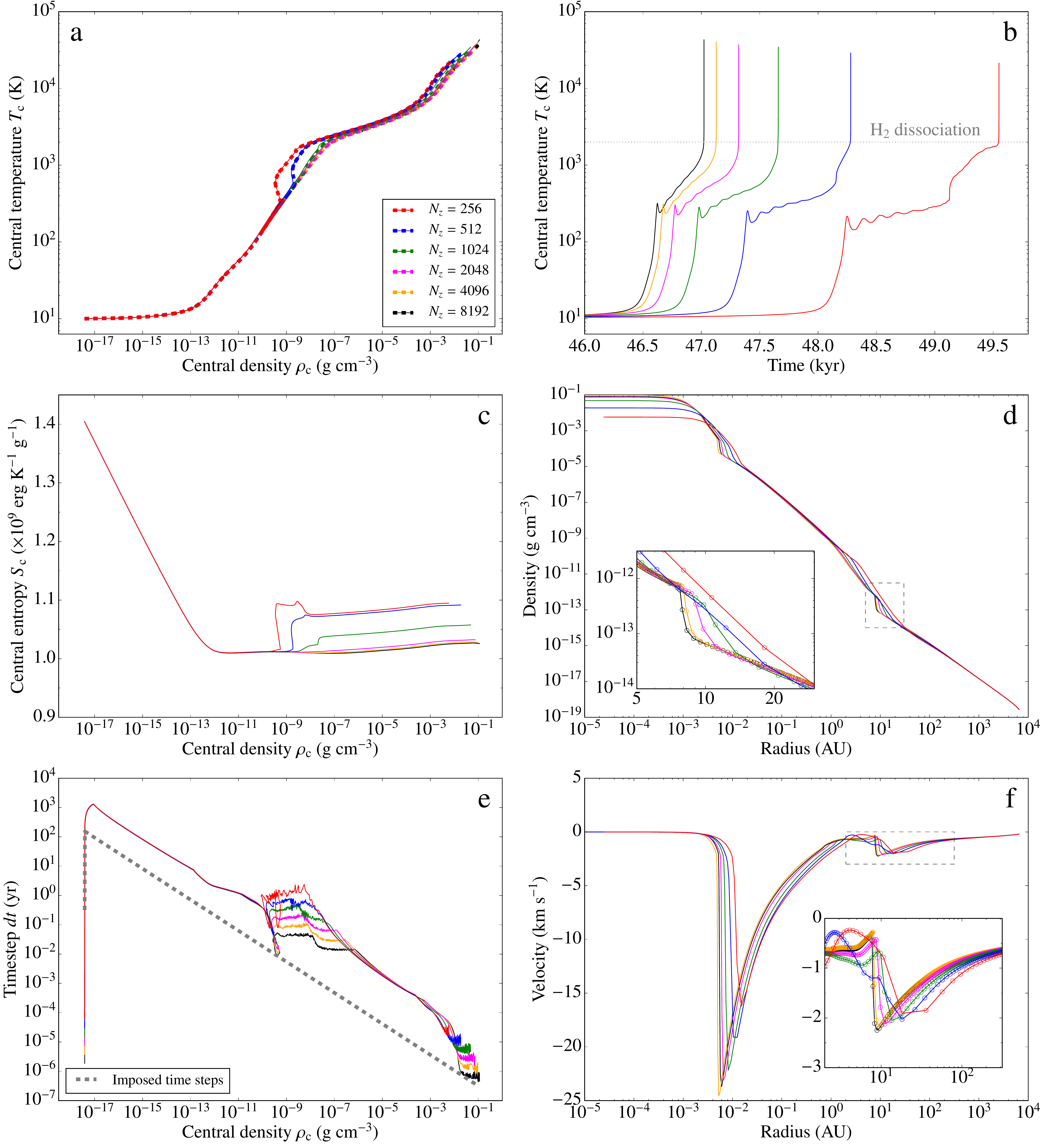}
\caption{Resolution study for run 69. Panels (a) and (c) show the temperature and entropy, respectively, of the cell at the centre of the system as a function of density 
using 256, 512, 1024, 2048, 4096, and 8192 cells (see legend). In panel (a), the thin solid lines are for simulations with the standard timestep calculation based on relative variations of primary variables from one time step to the next, while the thick dashed lines are for the re-runs when the values of $dt$ are enforced according to the central gas density (see text).
Panel (b) shows the central temperature as a function of time. 
Panels (d) and (f) show profiles of the gas density and velocity, respectively, as a function of radius, at the end of the simulations. Panel (e) shows the integration 
timesteps in the different simulations as a function of central density. The thick grey dashed line represents the forced values of $dt$ we have chosen for our temporal 
resolution re-runs (see text).}
\label{fig:resolution}
\end{figure*}


\subsection{Spatial resolution}

A resolution study was carried out to ensure the results presented in this work are numerically converged. Indeed, the evolution of the first and second Larson cores are 
very sensitive to numerical resolution, as enough cells are needed to adequately sample not only the Jeans length but also the accretion shocks at the core borders. A 
typical run was chosen (run 69) and re-simulated at different resolutions. The same scheme to increase grid cell size with increasing radius was used in all runs, but to allow for a fair comparison, 
the value of $\alpha$ was changed for each resolution. Indeed, for 4096 cells, the size ratio between the innermost and outermost cells is $1.0008^{4095} \simeq 26.4$. 
If we used the same $\alpha$ for 256 cells, this ratio would stand at $1.0008^{255} \simeq 1.23$. By considering only two cells in the grid, the first cell has size $\Delta r$ and the second has size $(1+\alpha_{0}) \Delta r$. If one wishes to double the resolution, i.e.~use 4 cells for the same mesh size and a constant size ration $(1+\alpha_{1})$, each coarse cell is split into two small cells. Let $\delta_{i} = 1 + \alpha_{i}$, where $i = 0,1$. This gives us the relations
\begin{equation}
\begin{array}{rcl}
dr + \delta_{1}dr & = & \Delta r\\
\delta_{1}^{2}dr + \delta_{1}^{3}dr & = & r_{0}\Delta r
\end{array}
\end{equation}
where $dr$ is the size of the first cell in the finer resolution mesh. This yields $\delta_{1} = \sqrt{r_{0}}$, or $\alpha_{1} = \sqrt{1 + \alpha_{0}} - 1$. For our 
resolution study, we adopt the final relation
\begin{equation}
\alpha_{N_{z}} = (1 + 0.0008)^{\frac{4096}{N_{z}}} - 1 ~,
\end{equation}
where $N_{z}$ is the number of grid cells. It is now easy to see that the size ration between the innermost and outermost cells
\begin{equation}
\frac{dr_{N_{z}}}{dr_{1}} = 1.0008^{4096 \left(\frac{N_{z}-1}{N_{z}}\right)}
\end{equation}
is almost independent of $N_{z}$, as long as the number of cells is large enough.

\begin{figure*}
\centering
\includegraphics[width=\textwidth]{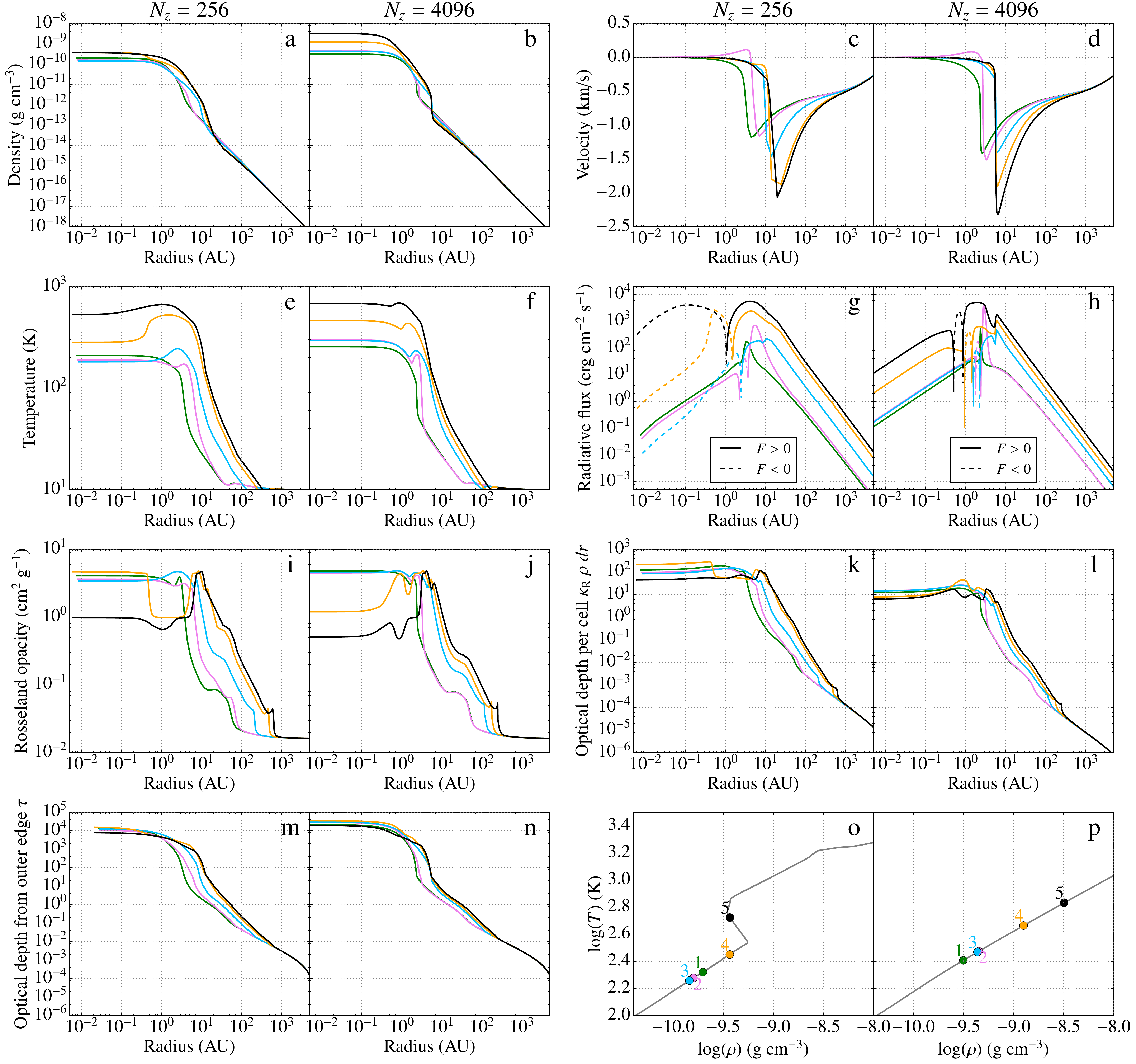}
\caption{Evolution sequence for two simulations with different resolutions. The first and third columns are for 256 grid cells, while 4096 cells have been used in the second and fourth columns. Panels (a) to (m) show radial profiles of various quantities, while panels (o) and (p) show a small portion of the thermal evolutions from Fig.~\ref{fig:resolution}a. The different colours represent a succession of epochs to illusrate the evolution of the system around the `knee' which appears in low-resolution runs. The numbers in panels (o) and (p) show the order in time of the colours. In panels (g) and (h), dashed lines represent negative values of the radiative flux (i.e.~a flux pointing towards the centre of the system).}
\label{fig:sequence}
\end{figure*}

The thermal evolutions at the centre of the system (temperature as a function of density) are shown in Fig.~\ref{fig:resolution}a (thin solid lines). While the first core is accreting, for 
central densities (in $\text{g~cm}^{-3}$) $10^{-10} \lesssim \rho_{c} \lesssim 10^{-8}$, the low resolution runs display a strong deviation from adiabaticity (we will refer to this as a `knee' below). During the 
second collapse ($10^{-8} \lesssim \rho_{c} \lesssim 10^{-4}$), all runs follow very similar adiabats, but the spread is again large once the second core has formed. 
This causes a significant spread in second core entropy in panel (c), and it is thus of prime importance that the simulations have the necessary resolution. The results 
appear to converge when 2048 cells and above are used.

Because the simulations are stopped as soon as a `hydrostatic bounce' occurs inside the second core (as soon as the density at the centre of the system decreases from 
one timestep to the next), the resolution also affects the final central densities and temperatures in the different runs. Table~\ref{tab:resolution} lists the final 
density, temperature and entropy at the centre of the system, for the different resolutions. It also lists in the last column the first core lifetime. This now suggests 
that a minimum resolution of 4096 is probably needed. This is confirmed by the temporal evolution of the central temperature in
Fig~\ref{fig:resolution}b. All simulations show a first hydrostatic bounce when the first core is formed
($T_{\text{c}} \sim 200-300$ K), but it occurs at different times. Simulations with lower resolutions then display several more bounces 
before proceeding to the second collapse, while the most resolved cases reach it in a much more `direct' manner. They also get there much faster, as illustrated by the 
wide spread in first core lifetimes $\tau_{\text{fc}}$ (i.e.~the time elapsed between the formation of the first and second cores) in the final column of Table~\ref{tab:resolution}. Finally, in panels (d) and (f), density and velocity profiles, respectively, show how well the accretion shocks at the first and second core 
borders are resolved.
Not only is it important to have an accurate description of the evolution of the gas at the centre of the system, it is also necessary to retain
a good enough resolution at the first and second core borders where the inflowing gas passes through sharp accretion shocks where density and temperature rise very abruptly.

The origin of the `knee' seen in Fig.~\ref{fig:resolution}a can be understood from looking at a succession of profiles, in two runs with different resolutions. This is 
shown in Fig.~\ref{fig:sequence} for 256 and 4096 cells, where different snapshots in time are represented by different colours. Since runs with different resolutions 
collapse and evolve at slightly different rates (see Fig.~\ref{fig:resolution}b for example), it does not make sense to compare profiles at exactly the same times, 
central temperatures or central densities. Instead, we have tried to select snapshots where the systems are approximately in the same states. The first snapshot 
(green) is at a time just when the first core accretion shock has formed. The second (pink) is when the gas velocity just behind the shock becomes positive and the core 
starts to expand. The third (blue) was taken when the core has settled and the gas velocities are once again negative everywhere in the grid. The final two 
snapshots (orange then black) show the subsequent evolution of the system, at times just before and just after the `knee' in the low-resolution run. We can see in panels 
(a,b,e,f,o,p) that at the time of first core formation, the high-resolution run reaches higher densities and temperatures in the centre. Nevertheless, the different 
velocity profiles are very similar (panels c \& d). We note that shortly after the formation of the first core, both high and low-resolution simulations show a small 
bump in gas temperature just behind the accretion shock, around 2-4 AU (best visible in the pink curve). The optical depth in both cases is well above unity (panels m \& 
n) and one can easily approximate the radiative flux to always be pointing in the direction of a temperature gradient\footnote{In a diffusion regime, it should 
technically be a gradient of radiative energy density, but radiation and gas are perfectly coupled inside the core, implying that radiative and gas temperature are 
equal.}. This can indeed be observed in panels (g) \& (h), where the radiative flux becomes negative between the summit of the temperature bump and the temperature 
minimum just to the left of the bump ($r\sim$ 2 AU)\footnote{This is a little hard to make out on the high resolution figure (panel h), but the sharp drops in the 
pink curve are a good indicator of a sign inversion.}. What happens next is interesting. Because the resolution is lower in the $N_{z}^{256}$ simulation, the optical 
depth per cell is higher than in the $N_{z}^{4096}$ run by about an order of magnitude (panels k \& l). The heating from the core gets trapped and the temperature in the 
bump rises. Contrastingly, in the high-resolution run, the heat is mostly radiated ahead of the shock; we can indeed see in panel (f) that the bump has hardly changed 
temperature between the pink and blue curves, but a sizeable radiative precursor has developed ahead of the shock (the temperature has risen by about a factor 5 between 
3 and 10 AU). Because the bump has not risen in temperature in run $N_{z}^{4096}$, the temperature in the centre of the system is still higher than the peak of the bump, 
and the radiative flux only shows a small sign inversion between 1.5-2.5 AU. In the $N_{z}^{256}$ calculation, the bump temperature exceeds the temperature at the centre 
of the core, and the radiative flux is now negative throughout almost the entire core (i.e.~to the left of the accretion shock; dashed lines in panel g). From here on, the flux continuously travels inwards, and begins to heat the core, from outside 
towards the centre. Indeed, in the orange curve of panel (e), we can see that the bump widens mostly in the inner direction. This heating Marshak wave travels 
towards smaller radii and eventually hits the centre of the system where the temperature rises abruptly. This causes the `knee' in panel (o). The last snapshot (black) 
shows profiles just at the end of the inward heating phase, right before the temperature profile inside the first core becomes flat again. In the case of 4096 grid 
points, the radiative flux is never fully directed towards the centre, and the temperature at the centre rises smoothly as time progresses. This shows that it is 
essential to resolve the characteristic lengths of all the radiative processes in our system. Typical radiation schemes usually require the optical depth per cell to remain below  
unity, but the asymptotic preserving scheme of \citet{Berthon2007} used in our code allows us to go to values ten times higher. The above analysis however shows that the 
solver cannot go much higher than $\kappa_{\mathrm{R}}\rho dr \sim 10$.

We have shown that the results converge for 4096 cells and above, and all 143 runs in the present work were thus run using 4096 cells. We are aware that this may change for a different set 
of initial conditions. We have carried out our resolutions study on a couple of other runs and found that the results were robust for $N_{z} = 4096$. Even though we have 
not done this for all the 143 runs, we are confident that the conclusions reached in this work are not affected by resolution effects.

\begin{table}
\centering
\caption{Density $\rho_{\text{c}}$, temperature $T_{\text{c}}$, and entropy $S_{\text{c}}$ at the centre of the system, at the end of the simulations, as a function of resolution. The last column lists the first core lifetime $\tau_{\text{fc}}$ (i.e.~the time elapsed between the formation of the first and second cores).}\label{tab:resolution}
\begin{tabular}{ccccc}
\hline
Resolution & $\rho_{\text{c}}$ & $T_{\text{c}}$ & $S_{\text{c}}$                      & $\tau_{\text{fc}}$\\
$N_{z}$    & (g cm$^{-3}$)     & (K)            & ($\text{erg~K}^{-1}~\text{g}^{-1}$) & (yr) \\
\hline
 256 & $5.66 \times 10^{-3}$ & 21194 & $1.095 \times 10^{-9}$ & 1302 \\
 512 & $1.83 \times 10^{-2}$ & 28732 & $1.092 \times 10^{-9}$ & 886  \\
1024 & $4.70 \times 10^{-2}$ & 34310 & $1.058 \times 10^{-9}$ & 679  \\
2048 & $7.35 \times 10^{-2}$ & 36635 & $1.032 \times 10^{-9}$ & 545  \\
4096 & $9.41 \times 10^{-2}$ & 39863 & $1.028 \times 10^{-9}$ & 455  \\
8192 & $1.11 \times 10^{-1}$ & 42673 & $1.026 \times 10^{-9}$ & 400  \\
\hline
\end{tabular}
\end{table}


\subsection{Temporal resolution}

In our time-implicit scheme, we limit the integration timesteps by constraining the variations from one time step to the next of each primary variable to a maximum of 10\% (see 
Sect.~\ref{sec:num}). Using a finer mesh also implies that smaller timesteps are used, and we may be seeing a double effect in the spatial resolution study above, as the 
lower spatial resolution runs may in fact also suffer from a lower temporal resolution. To verify this, we have re-run the simulations by forcing the timesteps to values 
similar to the highest resolution run (8192 cells). The timesteps of the above runs as a function of central density are plotted in Fig.~\ref{fig:resolution}e. The 
timesteps in all simulations follow a similar trend. They start at a very low value, set by the radiation CFL which defines the initial timestep to avoid any convergence 
issues at the beginning of the runs. They rise sharply until they reach about 1000 yr, where they become limited by strong variations of primary 
variables. The timesteps gradually fall as the system collapses. We chose a piecewise law for the value of $dt$ as a function of central density which is represented by 
the thick grey dashed line in Fig.~\ref{fig:resolution}e, which remains below all the curves at all times.
The results of the re-runs with forced timesteps are shown in Fig.~\ref{fig:resolution}a (thick dashed lines). We see that the thermal evolutions are virtually identical to the original runs, and a higher temporal resolution does not alleviate the departures from adiabaticity in low spatial resolution runs.


\section{Correlations weighting scheme}\label{app:corr_weighting}

\begin{figure*}
\centering
\includegraphics[width=\textwidth]{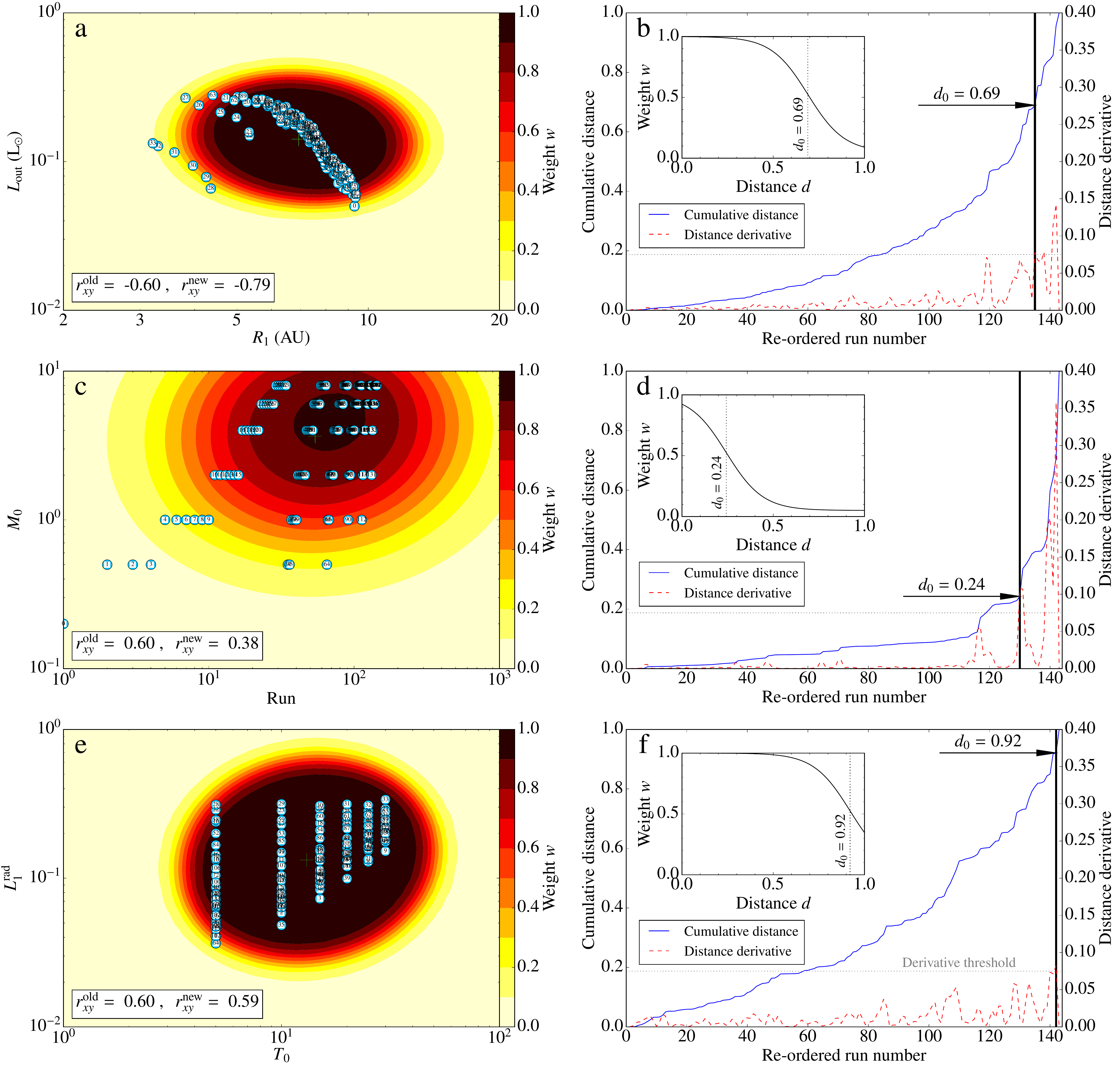}
\caption{Illustration of the weighting scheme for computing correlations between two variables $x$ and $y$. The left column shows the relation between $L_{\mathrm{out}}$ and $\dot{R}_{1}$, while the right is for $L_{\mathrm{out}}$ as a function of  $R_{0}$. In the top row, the blue dots represent the different simulations (the run numbers are indicated inside each symbol), while the coloured contours show the normalised weights given to the data points. In the bottom row, the blue solid line represents the cumulative distances for each run, sorted in ascending order. The red dashed line is the derivative of this distance distribution. The thick vertical black solid line marks the position of the first derivative spike above 0.05 (grey dotted line), while the black arrow reads the value of the distance $d_{0}$ on the blue curve at the location of the derivative spike. The insets show the weighting functions $w(d,d_{0})$.}
\label{fig:weighting}
\end{figure*}

When computing simple linear correlations between two quantities $x$ and $y$ where all the points were given the same weight of 1, we found that several correlation 
coefficients $r_{xy}$ were poorly reflecting $(x,y)$ relations which appeared quite strong, apart from a few outliers. Others seemed too high for variables which did not seem connected in any way, and outliers were setting the value of $r_{xy}$. This is illustrated in Fig.~\ref{fig:weighting}, left column, 
where three relations with identical unweighted correlations $r_{xy}^{\mathrm{old}} = 0.6$ are displayed. It is evident from the figure that the relation in the top panel shows in fact 
quite a tight correlation between the outer luminosity $L_{\mathrm{out}}$ and the first core radius $R_{1}$, with a few outliers, while the middle and bottom 
correlations, are relatively weak. The outliers have a strong influence on the final result, and it is well known that visual inspection is usually a mandatory check when working with linear correlations. It is never trivial to robustly identify outliers from a mathematical point of view, while they are in practise quite easy to spot by eye. To avoid having to inspect the 462 correlation plots, we devised a scheme to reduce the influence of the outliers on $r_{xy}$.

Outliers can be defined as isolated points that are distant from the majority of the other data points, a property that we shall take advantage of in our weighting scheme. To determine if a point is distant from the rest of the group, we computed the cumulative distance of a given point $i$ with all the other points $j$. We have
\begin{equation}
D_{i} = \sum_{j} \sqrt{\left(\frac{x_{j}-x_{i}}{x_{\mathrm{max}}-x_{\mathrm{min}}}\right)^{2} + \left(\frac{y_{j}-y_{i}}{y_{\mathrm{max}}-y_{\mathrm{min}}}\right)^{2}} 
\end{equation}
and
\begin{equation}
d_{i} = \frac{D_{i} - D_{\mathrm{min}}}{D_{\mathrm{max}} - D_{\mathrm{min}}} ~,
\end{equation}
where the min and max values are used to normalise the distances. If a point lies in the middle of the cluster, many of its distances to other points will be rather small, and only a few large distances towards the outliers will be recorded. In the case of an outlier, almost all of the distances computed will be large, amounting to a very high cumulative distance. We now need to choose a weighting function which gives a large weight to points with small distances and vice-versa. Many options exist, and common choices often involve $1/r$ or $1/r^{2}$ laws. We have found these to decrease too fast, giving too much weight to the points lying in the middle of a cluster, yielding too high correlation values computed from only a few data points from the sample. An optimal function should have an almost constant weight of 1 for most of the cluster, and a relatively sharp drop when distances become too large.
The function
\begin{equation}
w(d) = \frac{1-b}{\exp(a(d-d_{0}))+1} + b
\end{equation}
has the required properties. The value of $a$ determines the steepness of the decrease, while the value of $d_{0}$ dictates the value of $d$ for which $w$ starts to drop 
towards zero. The additional constant $0 < b < 1$ makes sure that no weight ever goes to zero to avoid discarding any data point completely. The arbitrary choices of $a=10$ and $b=0.05$ were made after some trial and error on several correlation relations. The value of $d_{0}$ however needs to be different for 
each correlation plot, as choosing a canonical value of 0.5 may cause many points to be omitted in both a correlated and non-correlated samples. For instance, in a 
perfectly correlated dataset, where all the points lie along the same line, the points in the middle of the line will have maximum distances of half the line length, 
while points at the extremities will have maximum distances of the entire line length, and consequently a larger cumulated distance. These points will be given a very 
low weight even though they are perfectly aligned with the others. This may not change the overall value of $r_{xy}$ in this idealised case, but it does limit the extent 
of the region of the parameter space over which $r_{xy}$ is determined. In a slightly less idealised case, where some scatter is present in the data, points far away 
from the mean of the sample but still lying along the same line may in fact help strengthen the correlation, or at least increase the accuracy of $r_{xy}$. On the other 
hand, in the case of a fully uncorrelated sample such as a random circular cloud of points, points towards the edges of the circle will have a much lower weight than 
the ones close to the centre, for no particular reason. By choosing an arbitrary value for $d_{0}$, one can strongly influence the statistical results in a biased way.

The method we have devised to objectively determine the value of $d_{0}$ for each plot is illustrated in the right column of Fig.~\ref{fig:weighting}. We first computed 
the cumulative distances for all the points in the sample, and sorted them in ascending order; this is represented by the blue solid line in panels (c) and (d). If one 
or more outliers are present in the data, there will be a sharp rise in the blue curve, as $d$ jumps from being in the cluster to lying at a large distance from the majority of the other 
points. The derivative of the sorted distance array is depicted by the red dashed line. We select the first derivative spike above 0.075 as the limit beyond which we are dealing with 
outliers (thick vertical black line). By then reading off the cumulative distance of the run responsible for the spike, we finally obtain our value for $d_{0}$. The 
small insets in panels (b), (d), and (f) show the weighting function for three different values of $d_{0}$.

We plot the 2D weighting function in the left panels using contours. In the top row, the weight drops moderately as one moves away from the 
cluster ($d_{0} = 0.69$), reducing the influence of the outliers in the bottom left corner of the plot. The value of $|r_{xy}|$ has increased from 0.60 to 0.79, now 
revealing the apparent correlation in the data which was not obvious in the non-weighted (`old') scheme.
In the middle row, the weight falls much more rapidly with increasing $d$ ($d_{0} = 0.24$) and the derivative quickly spikes above the threshold value of 0.075. This limits the effects of the outliers in the bottom left corner, lowering the value of $|r_{xy}|$ from 0.60 to 0.38.
In the bottom row, as the points are pretty much evenly spread out 
over the viewport, all the cumulative distances are similar, and the derivatives only exceed the threshold for the last few data points. Consequently, the value of $d_{0} = 0.92$ is large, most of the data points have a weight close to 1, and the value of $r_{xy}$ is virtually unchanged. We have inspected the vast majority of our final correlation plots by eye to make sure the weighting scheme was robust and produced the 
desired results.


\newpage

\section{List of run parameters}\label{app:paramstable}

Table~\ref{tab:parameters} lists the initial parameters and the properties of the first and second Larson cores for all 143 runs. The columns are (from left to right):

\bgroup
\def\arraystretch{1.4}
\noindent
\begin{tabular}{@{}l@{~~~}l@{~~~~}l@{}}
 1. &                                                                          & Run number                               \\
 2. & $M_{0}$                                                                  & Initial cloud mass                       \\
 3. & $T_{0}$                                                                  & Initial cloud temperature                \\
 4. & $R_{0} $                                                                 & Initial cloud radius                     \\
 5. & $\rho_{\mathrm{c}} = \frac{3M_{0}}{4\pi R_{0}^{3}}$                               & Initial cloud central gas density        \\
 6. & $\epsilon = M_{\text{BE}}/M_{0}$                                         & Bonnor-Ebert mass over initial mass      \\
 7. & $t_{\text{ff1}} = \sqrt{ \frac{3\pi}{32 G \rho_{\text{c}}} }$            & Free-fall time 1                         \\
 8. & $t_{\text{ff2}} = \frac{\pi}{2} \sqrt{ \frac{R_{0}^{3}}{2 G M_{0}} }$    & Free-fall time 2                         \\
\end{tabular}
\egroup

~\\
In the above, the quantities with a star subscript ($\star$) are evaluated at the relevant core border (first or second core), and $F$ represents the radiative flux. The simulation end time for each run is equivalent to $t_{\text{ff3}}$ (see Sect.~\ref{sec:tff_tend}).

\onecolumn
\footnotesize
\rowcolors{2}{gray!20}{white}
\begin{landscape}
\begin{center}
\begin{longtable}{c@{~~~}c@{~~~}c@{~~~}c@{~~~}c@{~~~}c@{~~~}c@{~~~}c@{~~~}c@{~~~}c@{~~~}c@{~~~}c@{~~~}c@{~~~}c@{~~~}c@{~~~}c@{~~~}c@{~~~}c@{~~~}c@{~~~}c@{~~~}c@{~~~}c}
\caption{Simulation initial parameters and core properties for all 143 runs.}\label{tab:parameters}\\

\hline
\hline
\rowcolor{white} Run        & $M_{0}$          & $T_{0}$ & $R_{0}$ & $\rho_{\mathrm{c}}$             & $\epsilon$ & $t_{\text{ff1}}$ & $t_{\text{ff2}}$ & $t_{\text{ff3}}$ & $t_{\text{sound}}$ & $R_{1}$ & $M_{1}$       & $\dot{M}_{1}$          & $L^{\text{acc}}_{1}$ & $L^{\text{rad}}_{1}$ & $\tau_{\text{fc}}$ & $R_{2}$ & $M_{2}$ & $\dot{M}_{2}$ & $L^{\text{acc}}_{2}$ & $L^{\text{rad}}_{2}$ & $L_{\text{out}}$\\
\rowcolor{white} ~          & (M$_{\odot}$)    &  (K)    & (kAU)   & (g cm$^{-3}$)            & ~           & (kyr)             & (kyr)             & (kyr)             & (kyr)          & (AU)    & (M$_{\odot}$) & (M$_{\odot}$/yr) & (L$_{\odot}$)        & (L$_{\odot}$)       & (yr) & (AU)    & (M$_{\odot}$) & (M$_{\odot}$/yr) & (L$_{\odot}$)        & (L$_{\odot}$) & (L$_{\odot}$)\\
\hline
\endfirsthead

\hline
\rowcolor{white} Run        & $M_{0}$          & $T_{0}$ & $R_{0}$ & $\rho_{\mathrm{c}}$             & $\epsilon$ & $t_{\text{ff1}}$ & $t_{\text{ff2}}$ & $t_{\text{ff3}}$ & $t_{\text{sound}}$ & $R_{1}$ & $M_{1}$       & $\dot{M}_{1}$          & $L^{\text{acc}}_{1}$ & $L^{\text{rad}}_{1}$ & $\tau_{\text{fc}}$ & $R_{2}$ & $M_{2}$ & $\dot{M}_{2}$ & $L^{\text{acc}}_{2}$ & $L^{\text{rad}}_{2}$ & $L_{\text{out}}$\\
\hline
\endhead

\hline
\endfoot

\hline \hline
\endlastfoot
 \href{http://starformation.hpc.ku.dk/images/collapse1d/display\_run.html?RunNumber=1}{1}  &       0.2  &         5  &         3  &    5.96e-18  &       0.74  &        27  &        65  &        54  &       106  &  9.30  &  3.10e-02  &  1.24e-05  &  6.02e-03  &  4.06e-02  &      1528 &  3.83e-03  &  2.60e-03  &  6.45e-02  &  6.41e+03  &  2.48e-09 &  5.00e-02\\ 
 \href{http://starformation.hpc.ku.dk/images/collapse1d/display\_run.html?RunNumber=2}{2}  &       0.5  &         5  &         3  &    1.49e-17  &       0.30  &        17  &        41  &        21  &       106  &  8.75  &  3.42e-02  &  2.44e-05  &  1.39e-02  &  5.19e-02  &       780 &  4.87e-03  &  2.58e-03  &  8.67e-02  &  6.71e+03  &  5.09e-09 &  6.64e-02\\ 
 \href{http://starformation.hpc.ku.dk/images/collapse1d/display\_run.html?RunNumber=3}{3}  &       0.5  &        10  &         3  &    1.49e-17  &       0.60  &        17  &        41  &        27  &        75  &  8.52  &  3.50e-02  &  3.12e-05  &  1.87e-02  &  5.83e-02  &       638 &  4.77e-03  &  2.60e-03  &  8.65e-02  &  6.89e+03  &  4.52e-09 &  7.15e-02\\ 
 \href{http://starformation.hpc.ku.dk/images/collapse1d/display\_run.html?RunNumber=4}{4}  &       0.5  &        15  &         3  &    1.49e-17  &       0.89  &        17  &        41  &        53  &        61  &  8.33  &  3.64e-02  &  4.63e-05  &  2.96e-02  &  7.28e-02  &       476 &  4.16e-03  &  2.64e-03  &  7.24e-02  &  6.71e+03  &  3.11e-09 &  8.22e-02\\ 
 \href{http://starformation.hpc.ku.dk/images/collapse1d/display\_run.html?RunNumber=5}{5}  &       1.0  &         5  &         3  &    2.98e-17  &       0.15  &        12  &        29  &        13  &       106  &  8.03  &  3.78e-02  &  5.40e-05  &  3.72e-02  &  7.78e-02  &       455 &  4.97e-03  &  2.58e-03  &  8.39e-02  &  6.35e+03  &  5.76e-09 &  8.98e-02\\
\hline
\end{longtable}
\end{center}
\end{landscape}

\end{document}